\documentstyle[12pt,epsf]{article}
\newcommand{\la}[1]{\label{#1}}

\newcommand{\ncs}{N_{\rm CS}}
\newcommand{\nc}{\newcommand}
\nc{\beq}{\begin{equation}}
\nc{\eeq}{\end{equation}}
\nc{\beqa}{\begin{eqnarray}}
\nc{\eeqa}{\end{eqnarray}}
\nc{\lra}{\leftrightarrow}
\nc{\lmax}{l_{\rm max}}

\nc{\sss}{\scriptscriptstyle}

\renewcommand{\vec}[1]{{\bf #1}}

\newcommand{\be}{\begin{equation}}
\newcommand{\ee}{\end{equation}}
\newcommand{\ba}{\begin{eqnarray}}
\newcommand{\ea}{\end{eqnarray}}
\newcommand{\bi}{\begin{itemize}}
\newcommand{\ei}{\end{itemize}}
\newcommand{\fig}{Fig.~}
\newcommand{\nr}[1]{(\ref{#1})}
\newcommand{\tr}{{\rm Tr\,}}

\newcommand{\fr}[2]{{\frac{#1}{#2}}}

\renewcommand{\vec}[1]{{\bf #1}}

\newcommand{\eq}{Eq.\,}
\newcommand{\eqs}{Eqs.\,}
\newcommand{\h}{\hspace*{4mm}}
\newcommand{\vk}{{\bf k}}
\newcommand{\vv}{{\bf v}}
\newcommand{\dd}{{\rm d}}
\newcommand{\cf}{c.f.\@}
\newcommand{\su}[1]{\mbox{SU(#1)}}
\newcommand{\smallfr}[2]{{\textstyle\fr{#1}{#2}}}
\newcommand{\half}{\smallfr12}

\newcommand{\dt}{\delta_t}
\newcommand{\trans}{{\cal P}}
\newcommand{\betaL}{\beta_{\rm L}}

\def\lsi{\raise0.3ex\hbox{$<$\kern-0.75em\raise-1.1ex\hbox{$\sim$}}}
\def\gsi{\raise0.3ex\hbox{$>$\kern-0.75em\raise-1.1ex\hbox{$\sim$}}}
\newcommand{\lsim}{\mathop{\lsi}}
\newcommand{\gsim}{\mathop{\gsi}}
\makeatletter \@addtoreset{equation}{section} \makeatother
\renewcommand{\theequation}{\arabic{section}.\arabic{equation}}

\begin{document}

\begin{titlepage}
\begin{flushright}
NORDITA-99/45HE\\
MCGILL-99/22\\
NBI-HE-99-25\\
hep-ph/9907545\\
\end{flushright}
\begin{centering}
\vfill

{\bf CHERN-SIMONS NUMBER DIFFUSION AND HARD THERMAL LOOPS
     ON THE LATTICE}
\vspace{0.8cm}

D. B\"odeker$^{\rm a,}$\footnote{bodeker@nbi.dk},
Guy D. Moore$^{\rm b,}$\footnote{guymoore@hep.physics.mcgill.ca} and
K. Rummukainen$^{\rm c,d,}$\footnote{kari@nordita.dk}

\vspace{0.3cm}
{\em $^{\rm a}$Niels Bohr Institute, Blegdamsvej 17, DK-2100 Copenhagen \O,
Denmark\\ }
\vspace{0.3cm}
{\em $^{\rm b}$Dept. of Physics, McGill University, 3600 University St.,
	Montreal, PQ H3A 2T8 Canada\\ }
\vspace{0.3cm}
{\em $^{\rm c}$Nordita, Blegdamsvej 17, DK-2100 Copenhagen \O,
Denmark\\ }
\vspace{0.3cm}
{\em $^{\rm d}$Helsinki Institute of Physics,
P.O.Box 9, 00014 University of Helsinki, Finland\\ }

\vspace{0.7cm}
{\bf Abstract}

\end{centering}

\vspace{0.3cm}\noindent
We develop a discrete lattice implementation of the hard thermal loop
effective action by the method of added auxiliary fields.  We use the
resulting model to measure the sphaleron rate (topological
susceptibility) of Yang-Mills theory at weak coupling.  Our results give
parametric behavior in accord with the arguments of Arnold, Son, and
Yaffe, and are in quantitative agreement with the results of Moore, Hu,
and M\"{u}ller.

\vfill
\vfill
\vfill
\noindent
PACS: 11.10.Wx, 11.15.Ha, 12.60.Jv, 98.80.Cq \\ Keywords: \\ finite
temperature, baryon number violation, electroweak phase transition,
lattice simulations

\vfill

\end{titlepage}

\section{Introduction}

Baryon number is not a conserved quantity in the standard model.
Rather, because of the anomaly, its violation is related to the
electromagnetic field strength of the weak SU(2) 
group \cite{tHooft},
\beq
\partial_\mu J^\mu_B = N_G \frac{g^2}{32 \pi^2} 
	\epsilon_{\mu \nu \alpha \beta} {\rm Tr} F^{\mu \nu} 
	F^{\alpha \beta} = N_G \frac{g^2}{8 \pi^2} E_i^a B_i^a \, ,
\la{fftilde}
\eeq
where $N_G=3$ is the number of generations.\footnote{There is also a
contribution from the hypercharge fields, but it will not be relevant
here because the topological structure of the abelian vacuum does not
permit a permanent baryon number change.}
In vacuum the efficiency of baryon number violation through this
mechanism is totally negligible \cite{tHooft}, but at a sufficiently
high temperature this is no longer 
true \cite{Kuzmin,ArnoldMcLerran}.  This can
have very interesting cosmological significance, since it complicates
GUT baryogenesis mechanisms and opens the possibility of baryogenesis
from electroweak physics alone.  This motivates a careful
investigation of baryon number violation in the standard model at high
temperatures.  

The baryon number violation rate relevant in cosmological settings can
be related by a fluctuation dissipation relation
\cite{KhlebnikovShaposhnikov,Mottola,RubakovShaposhnikov} to the
``Minkowski topological susceptibility'' of the electroweak theory, 
also called the ``sphaleron rate,''
\beq
\Gamma \equiv \int d^3 x \int_{- \infty}^{\infty} dt
	\left( \frac{g^2}{8 \pi^2} \right)^2 \langle [E_i^a B_i^a(x,t)] 
	\: [E_j^b B_j^b(0,0)] \rangle \, ,
\la{Def-of-Gamma}
\eeq
where $\langle \rangle$ means an expectation value  with respect to the equilibrium
thermal density matrix.  Here $t$ is Minkowski time.  This quantity is
{\em not} simply related to the Euclidean topological susceptibility
\cite{ArnoldMcLerran2}, and we do not possess either perturbative tools
or Euclidean tools to carry out its calculation.

It has been argued by Grigoriev and Rubakov \cite{GrigRub} that
the value of the susceptibility $\Gamma$ in the quantum theory will be
the same as its value in classical Yang-Mills field theory.  This would
open a new avenue for 
measuring $\Gamma$, since classical Yang-Mills theory can be put on the
lattice \cite{Ambjornetal}.  There has been some progress on measuring
$\Gamma$ on the lattice \cite{AmbKras,Moore1,TangSmit,AmbKras2}; in
particular two different methods have been developed for dealing with
the right hand side of Eq. (\ref{Def-of-Gamma}) in a topological way
which eliminates lattice artifacts in its measurement
\cite{slavepaper,broken_nonpert}.  

At the same time our qualitative understanding of Grigoriev and Rubakov's
claim has improved.  A complication with their proposal is that 3+1
dimensional classical Yang-Mills theory contains ultraviolet (UV)
divergences, which B\"{o}deker, McLerran, and Smilga have argued may be
important in setting $\Gamma$ \cite{Smilga}.  Subsequently, Arnold, Son,
and Yaffe have demonstrated that a particular class of diagrams,
the hard thermal loops (HTL's), are essential to establishing $\Gamma$ 
\cite{ASY}.  The amplitude of the HTL's in the classical theory is
linearly divergent, and therefore linearly cutoff dependent.  In the
full quantum theory the HTL's are finite, with almost all of the
contribution arising from excitations with momentum $k \simeq \pi T$;
such high $k$ excitations 
are not properly described by the classical theory.  Arnold, Son,
and Yaffe argue that because of the HTL's, the effective infrared (IR)
theory ``feels'' the ``cutoff'' which quantum mechanics provides for the
classical theory, and that the value of $\Gamma$ scales inversely with
the cutoff momentum scale.  As a result, rather than the naive
dimensional estimate of $\Gamma \sim \alpha^4 T^4$, the parametric
behavior of $\Gamma$ should be $\Gamma \sim \alpha^5 T^4$; and in
particular $\Gamma$ is inversely proportional to the strength of the
HTL's, which is conveniently parameterized by the Debye mass squared
$m_{\rm D}^2$.  On the lattice this means that $\Gamma$ should depend on the
lattice spacing $a$ as $\Gamma \propto a g^2T \alpha^4 T^4$, a behavior
which has recently been verified numerically \cite{MooreRummukainen}. 
Arnold, Son, and Yaffe's argument has been carefully
re-analyzed by B\"{o}deker, 
who has shown that there is an additional, logarithmic dependence on the
Debye mass, and that, permitting an expansion in $\log(1/g) \gg 1$, the
leading behavior is actually $\Gamma \sim \alpha^5 \log(1/g) T^4$
\cite{Bodeker}.

If we take the limit $\log(1/g) \gg 1$, B\"{o}deker presents an
effective theory for evaluating the coefficient of the $\alpha^5
\log(1/g) T^4$ law \cite{Bodeker}.  The effective theory 
is UV safe \cite{ASY2} and the coefficient can
be found accurately by lattice means \cite{Bodek_paper}.
However, in practice the expansion in $\log(1/g) \gg 1$ turns out to be
very poorly behaved.  To get a reasonably accurate value for $\Gamma$ at
the physical value for the electroweak coupling, $\alpha \simeq 1/30$,
it is necessary to treat the dynamics of the classical field theory with
a full inclusion of the HTL effects.  This is challenging, because the
HTL effective action is nonlocal \cite{HTLaction}.  However, it is
possible to rewrite the HTL action in terms of a local theory with added
degrees of freedom, as we will discuss below.  Thus, it could be
possible to determine $\Gamma$ by measuring the topological
susceptibility of lattice regulated, classical Yang-Mills theory,
supplemented by added degrees of freedom which correctly generate the
hard thermal loop effects.  Doing so would both test Arnold, Son, and
Yaffe's claim, and determine the numerical coefficient of the $\alpha^5
T^4$ law, and therefore tell us how efficiently baryon number is
violated at high temperatures.

One way of realizing this goal was presented in \cite{HuMuller} and
implemented and used to measure $\Gamma$ in \cite{particles}.  
The purpose of this paper is to present
an alternative and in some respects more efficient implementation of
classical Yang-Mills theory plus hard thermal loops, and to use it to
check the results of \cite{particles}.  
Our approach is based on a way of writing the hard thermal loops in
terms of auxiliary fields which was first proposed in
\cite{BlaizotIancu1}.  Using this formulation to incorporate the HTL
action on the lattice has been advocated by B\"{o}deker, McLerran, and
Smilga \cite{Smilga}.  This
paper represents a concrete numerical realization of that idea.

In Section \ref{HTL-continuum} we review the local
formulation of classical Yang-Mills field theory 
supplemented by the HTL action  due to Blaizot and Iancu 
and due to Nair.  Their theory contains an infinite set
of fields, so in Section \ref{Harmonics-sec} we
perform a transformation and a truncation to make the number of fields
in the model finite, without losing spherical symmetry.  The resulting
theory does not quite give the correct HTL equations of motion; we study the
difference, and how it vanishes in the limit as the truncation leaves
in more and more fields, in Section \ref{Propagator-sec}.  Then we
discretize space and time in Section \ref{sec:lattice},
and review how to measure $\Gamma$ topologically in Section
\ref{topology}.  We study the numerical behavior of $\Gamma$ as
a function of the strength of the HTL's and the truncation point in
Section \ref{sec:results}.  

Our conclusions are in Section
\ref{Conclusion-sec}, but we summarize them here.  The HTL effective
theory shows a dependence on the strength of HTL's which is consistent
with Arnold, Son, and Yaffe's arguments, and grossly inconsistent with
HTL independence.  The dependence on the truncation point is
surprisingly weak, so only a few new fields need to be added to
approximate the correct HTL behavior.  Thus, our algorithm proves
quite an efficient way of incorporating HTL's.  Our final results for
$\Gamma$ are consistent with those of Moore, Hu, and M\"{u}ller
\cite{particles}, and for the physical value of $m_{\rm D}^2$ in the
minimal standard model, $m_{\rm D}^2 = (11/6) g^2 T^2$ and $\alpha =
1/30$, they give approximately $\Gamma = (25.4 \pm 2.0)\, \alpha^5 T^4$.

\section{Hard thermal loops in the continuum}
\la{HTL-continuum}

In this section we discuss the origin of the hard thermal loops in terms
of kinetic theory, and we present a local theory in which extra degrees of
freedom generate the hard thermal loops.  Nothing in this section is
original; rather it is a review of Blaizot and Iancu's and of Nair's work
\cite{BlaizotIancu1,Nair:local,Blaizot:energy,Nair:hamiltonian}.  
We include it for completeness and because our
numerical implementation of classical Yang-Mills theory with hard
thermal loops will be built directly from it.

Two controlled approximations make the dynamics of IR fields in the
electroweak theory tractable numerically, and both arise because the
theory is weakly coupled.  First, the IR degrees of freedom can to a
good approximation be treated as {\em classical} fields.  
Using this fact to perform calculations of nonperturbative IR
correlators was first proposed by Grigoriev and Rubakov \cite{GrigRub},
and the accuracy of the approximation has been addressed 
in \cite{Smilga,MooreTurok,Bodeker_classical}.  The conclusion of
\cite{Bodeker_classical} is that the classical approximation is an
excellent approximation in the infrared, but UV divergences in the
classical theory are potentially dangerous and must be handled
carefully. 

The solution to this problem is to regulate the classical theory in some
way, which for the moment we will not specify, and then to treat the UV
degrees of freedom separately by perturbation theory.  Here the other
controlled approximation enters; the UV degrees of freedom are described
by linearized kinetic theory, up to corrections subleading in $g$.

Since the equilibrium distribution of UV modes, $N_0(\vk)$, is color neutral,
it does not directly enter in the field equations of the classical IR
fields.  Rather, it is necessary to expand the UV mode distribution
function (one particle density matrix) up to first order in fluctuations
from equilibrium, 
\beq
N(x,\vk) = N_0(\vk) + \delta N_{\rm singlet}(x,\vk) + 
	\delta N_{\rm adj.}(x,\vk) 
	+ \ldots \, .
\eeq
Fields in a representation higher than fundamental lead, in addition to
the singlet and adjoint representation terms we have written, to higher
representation departures from equilibrium; but neither these, nor the
singlet deviation from 
equilibrium $\delta N_{\rm singlet}$, directly interact with the IR
classical fields, and at the linearized level they can be dropped; only
$N_0$ and $\delta N_{\rm adj.}$ will be relevant.  
Note also that $N$ should have
a spin index, and if there are scalar or fermionic degrees of freedom
then it also has a species index. At leading order, corresponding to the HTL 
approximation, the contribution
from each spin and species are of the same form except in the statistics
for $N_0$, so we will not write them in what follows. 

At leading order in the coupling the IR classical fields evolve under
the Yang-Mills field equations with a source arising from the UV modes,
\cite{BlaizotIancu1} 
\beqa
(D^\nu F_{\nu \mu})^a & = & j_\mu^a \, , \\
j_\mu^a (x)& = & 2 g C_A \int \frac{d^3 k}{(2 \pi)^3} v_\mu \delta N^a(x,\vk)
	\, , 
\la{YM-equation}
\eeqa
with $v^\mu = (1,\vv)$, $\vv = \vk/|\vk|$ the (ultrarelativistic)
3-velocity of the particles (note that $v^\mu$ is not a Lorentz
covariant quantity), and $C_A=2$ for SU(2) gauge theory.  We have only
written the contribution of gauge excitations here, there are
additional terms of the same form for scalars and fermions where
appropriate.  The distribution function evolves via a convective
covariant derivative equation which reflects the ultrarelativistic
propagation of the UV degrees of freedom.  The interactions between
$\delta N_a$ and the IR classical field strength is subdominant
because the coupling is weak; however, the electric field polarizes
the equilibrium distribution, providing a source term for $\delta
N_a$.  The equation for the evolution of $\delta N^a$, at leading
order in $g$, is
\beq
 \frac{\dd \delta N^a}{\dd t} = 
	(v_{\mu} D^{\mu}_x)^{ab} \delta N^b(x,\vk) + 
	g v_\mu F_{0\mu}^a(x) 
	\frac{\partial N_0}{\partial |\vk|} = 0\, .
\la{convective-N}
\eeq
Note that this equation is not Lorentz covariant; it involves only the
electric field, not the magnetic field.  The reason is that the
equilibrium distribution $N_0$ has a rest frame.  A magnetic field in
that frame changes trajectories of individual particles, but it does not
disturb the (rotationally symmetric) equilibrium distribution, whereas an
electric field polarizes the plasma.

One approach to making a numerical model for the IR classical fields
plus UV modes is to simulate the distribution function $N$ with a large
number of charged particle degrees of freedom.  In the limit that the
number of particles is large and their charges are small, one recovers
the above equations.  This is the approach proposed by \cite{HuMuller}
and implemented in \cite{particles}.  Here we will deal instead with the
distribution functions.  This complementary approach can test the
reliability of the results of \cite{particles} and may also prove
simpler and more efficient.  This is particularly true because
Eqs. (\ref{YM-equation}) and (\ref{convective-N}) carry extra redundant
information; $\delta N^a$ is actually a function of $\hat{\vk}$ times a
fixed function of $|\vk|$, namely, \cite{Nair:local,Blaizot:energy}
\beq
\delta N^a(x,\vk) = - g\frac{\partial N_0}{\partial |\vk|} W^a(x,\vv) \, , 
\eeq
where $\vv = \hat{\vk}$ takes on values over the unit sphere.  In terms of
$W$, the convective evolution of the departure from equilibrium is
\beq
(v_{\mu} D^{\mu})^{ab} W^b(x,\vv) = v^{\mu} F_{0 \mu}^a(x) \, ,
\la{W-evolution}
\eeq
and the current felt by the IR classical fields is
\beq
j_\mu^a (x)= m_{\rm D}^2 \int \frac{d \Omega_v}{4 \pi} v_\mu W^a(x,\vv) \, ,
\la{W-current}
\eeq
where $d\Omega_v$ means that $\vv$ is integrated over the unit sphere with
its natural measure.  Here $m_{\rm D}^2$ is the square of the Debye mass.
These equations can be viewed as generalized Hamilton-Jacobi equations
arising from the conserved energy\footnote{Throughout this paper Roman
direction indices run over the 3 spatial directions with positive
metric, while Greek direction indices run over all 4 spacetime indices
with signature ($+---$).}
\beq
H = \int d^3 x \left( \frac{1}{4} F_{ij}^a F_{ij}^a +
	\frac{1}{2} F_{01}^a F_{01}^a +  
	\half m_{\rm D}^2 \int \frac{d\Omega_v}{4 \pi} W^a(x,\vv) W^a(x,\vv) \right) 
	\, ,
\label{W-Hamiltonian}
\eeq
and rather nontrivial Lie-Poisson brackets 
\cite{Nair:hamiltonian}.  Here
$\Omega_v$ is the integration measure for integrating $\vv$ over the
sphere.  

When there are more than one species, $W^a$ represents the deviation
from equilibrium felt by each, and $m_{\rm D}^2$ is a sum of a contribution
from each species of charge carrier,
\beq
\label{mD}
m_{\rm D}^2 = g^2 T^2 \left( \frac{N}{3} + \frac{N_s}{6} + \frac{N_f}{12}
	\right) \, ,
\eeq
with $N_s$ the number of fundamental representation, complex scalars and
$N_f$ the number of fundamental representation, chiral fermions.  In the
SU(2) weak sector of the minimal standard model, $N=2$, $N_s = 1$, 
and $N_f = 12$, so $m_{\rm D}^2 = (11/6) g^2 T^2$.  This is also a lower bound
for all extensions of the standard model.

Our approach will be to find a discrete implementation of
Eqs. (\ref{YM-equation}), (\ref{W-evolution}), and (\ref{W-current}),
and to study their evolution to determine the diffusion constant for
Chern-Simons number.  

\section{Expansion in Spherical Harmonics}
\la{Harmonics-sec}

Unfortunately, the representation of the hard thermal loops in terms of
$W^a(x,\vv)$ does not provide a set of equations which are easy to implement
numerically.  The problem is that $W^a(x,\vv)$ is a function not only of 
space-time,
but also over the sphere.  Even if we discretize space onto a lattice,
$W$ still ``lives'' on a sphere at each lattice point, so it still takes
an infinite amount of information to specify $W$ completely.  It is
necessary to define $W$ over the sphere in some way requiring only a
finite number of degrees of freedom.  Since we want to recover spherical
symmetry on scales long compared to our lattice spacing, we should
choose to do so in a spherically symmetric way.  Our choice is to expand
$W$ in spherical harmonics,
\beq
W^a(x,\vv) = \sum_{l=0}^{\infty} \sum_{m=-l}^{l} 
	W_{lm}^a(x) Y_{lm}(\vv) \, ,
\la{W-in-Ylms}
\eeq
where $W_{lm}^a(x)$ is a function over space-time only.  Because $W^a(x,\vv)$
is real valued, the $W_{lm}^a$ satisfy the
relations 
\beq
W^a_{lm} = (-1)^m W^{a*}_{l,-m} \, ,
\eeq
so only the real part of $W_{l0}$, and the real and imaginary parts of 
$W_{lm}$ for $m>0$, should be viewed as independent variables.  Here
we use the Condon-Shortley phase convention and normalize $Y_{lm}$ 
so that 
\be
  \int {d\Omega}\, Y^*_{lm} Y_{l'm'} = \delta_{l,l'}\delta_{m,m'}\,.
\ee
Inserting the expansion (\ref{W-in-Ylms}) into Eq. (\ref{W-evolution}),
multiplying by $Y^*_{lm}$, and integrating over angles, gives the equation
of motion for $W_{lm}^a$, 
\beq
\frac{\partial W^a_{lm}}{\partial t} = -C_{lm,l'm',i}\,\, (D_i)^{ab} W^b_{l'm'}
	+ \delta_{l,1} v_{mi} E_i^a \, ,
\la{update-lm}
\eeq
where $v_{mi}$ is the vector $\vv$ expressed in spherical components
\be
v_{mi} = \int {d\Omega_v} Y^*_{1m}(\vv) v_i 
\la{dmi}
\ee
and $C_{lm,l'm',i}$ is an integral over 3 spherical harmonics:
\beq
C_{lm,l'm',i} = \int {d\Omega_v} Y^*_{lm}(\vv) v_i Y_{l'm'}(\vv) \,.
\la{clm}
\eeq
We give explicit expressions for $v_{mi}$ and $C_{lm,l'm',i}$ in
Appendix \ref{app:1}.

Furthermore, in terms of the spherical components the current is
\beq
j_i^a = \fr{m_{\rm D}^2}{4\pi} v^*_{mi} W_{1m}^a \, , \qquad
j_0^a = \fr{m_{\rm D}^2}{\sqrt{4\pi}} W_{00}^a \, ,
\la{current-lm}
\eeq
and the conserved energy density is
\beq
H = \int d^3 x \left( \frac{1}{2} ( E_i^a E_i^a + B_i^a B_i^a ) + 
	\fr12 \fr{m_{\rm D}^2}{4\pi} \sum_{lm} |W^a_{lm}|^2 \right) \, .
\la{Hamiltonian-lm}
\eeq
As written, Eqs. (\ref{update-lm}), (\ref{current-lm}), and
(\ref{Hamiltonian-lm}) are
equivalent to Eqs. (\ref{W-evolution}), (\ref{W-current}), and
(\ref{W-Hamiltonian}).  They
still contain an infinite number of degrees of freedom.  However, they
are in a form more amenable to a spherically symmetric truncation.

The meaning of $m_{\rm D}^2 W^a(x,\vv)$ is that it represents the net charge of
all excitations moving in the $\vv$ direction at point $x$.  What we have
done is to transform to angular moments.  $m_{\rm D}^2 W^a_{00}(x)$ is the total
charge of all excitations at site $x$.  $m_{\rm D}^2 W^a_{10}(x)$ is roughly the
net charge moving in the $+z$ direction minus charge moving in the $-z$
direction, and $m_{\rm D}^2 W^a_{lm}$ with $l \geq 2$ represent higher tensor
moments in the distribution of excitations.  For instance, a positive
$m_{\rm D}^2 W^a_{20}$ means, roughly, that there are more charges of type $a$
moving either up or down the $z$ axis than in the $x,y$ plane.
Note that only $W_{00}^a$ and $W_{1m}^a$ interact with the IR fields,
and only $W_{1m}^a$ is directly sourced by those fields.  All the higher
moments are important only in propagating the charge distribution
through the convective derivative term in \eq(\ref{update-lm}).

The model still contains a countably infinite number of degrees of freedom,
namely $W^a_{lm},$ $l=1,2,3,\ldots$ .  Most of these degrees of freedom
are describing extremely subtle, high tensor fluctuations in the
distribution of moving charges.  It is reasonable to think that smearing
the angular resolution of the distribution of charges by truncating the
series of $Y_{lm}$ at some finite $\lmax$ will not significantly change
the physics.  In particular, for $\lmax \geq 1$ it will not change the
way the charge current interacts with the Yang-Mills fields, but only
the way the charges propagate; and for sufficiently large $\lmax$ we
expect the effect of angular smearing to be unimportant.  Therefore, to
render the set of fields finite, we truncate the series of 
$W_{lm}^a$ at some finite $\lmax$.  The evolution equation for
$W_{lm}^a$ is still Eq. (\ref{update-lm}), but with all $W_{l'm'}^a$
with $l'>\lmax$ fixed to zero.  Equivalently, we could set all
$C_{lm,l'm',i}$ with either $l>\lmax$ or $l'>\lmax$ to zero.  The
number of independent adjoint $W_{lm}$ matrices is $(\lmax+1)^2$.

As long as $C_{lm,l'm',i}$ satisfies the relation
\beq
C_{lm,l'm',i} = C^*_{l'm',lm,i}
\eeq
and the terms involving $v_{mi}$ are either both present or both absent
(they are absent if $\lmax = 0$), then the Hamiltonian,
Eq. (\ref{Hamiltonian-lm}), and the phase space measure are conserved by
the evolution equations.
Hence it makes sense to speak of equal and unequal time, equilibrium
thermal correlation functions.  When $\lmax$ is finite we are no longer
considering a theory which is strictly equivalent to classical
Yang-Mills field theory with added hard thermal loops, but the behavior
should approach the correct behavior in the limit $\lmax \rightarrow
\infty$ and we can consider taking this limit numerically.

\section{Propagator and Thermodynamics at Finite $\lmax$}
\la{Propagator-sec}

Before moving on to the numerical implementation of the effective
theory described in the last section in discrete space, we should
pause to see how well or how badly the theory with finite $\lmax$
cutoff reproduces the hard thermal loops.  To do so we first look at
whether it reproduces them correctly at the thermodynamic level; it
does so perfectly for all $\lmax \geq 0$.  Then we examine the
propagator of the theory, which will only be reproduced properly in
the $\lmax
\rightarrow \infty$ limit.

\subsection{Thermodynamics}
\label{thermo-subsec}

As discussed at the end of the last section, the theory with an $\lmax$
cutoff possesses well defined thermodynamics described by a Hamiltonian
which is quadratic and diagonal in the $W_{lm}$'s.  The only complication
is that the phase space is constrained due to Gauss' law:
\beq
(D \cdot E)^a = \frac{m_{\rm D}^2}{\sqrt{4 \pi}} W^a_{00} \, ,
\eeq
and the partition function reads
\beqa
Z & = & \int {\cal D} A_i {\cal D} E_i {\cal D} W_{lm} \delta \left( 
	(D \cdot E)^a - m_{\rm D}^2 W^a_{00} / 
	\sqrt{4 \pi} \right) \exp(-H/T) \, , \\
H & = & \frac{1}{2} \int d^3 x \left( B^a_i B^a_i + E^a_i E^a_i + 
	\frac{m_{\rm D}^2}{4 \pi} \sum_{lm} |W^a_{lm}|^2 \right) \, .
\eeqa
Every $W_{lm}$ except $W_{00}$ is Gaussian and they can all be
integrated out immediately.  It is also convenient to introduce a
Lagrange multiplier for Gauss' law,
\beq
\delta\left( (D \cdot E)^a - m_{\rm D}^2 W^a_{00}/\sqrt{4 \pi}\right) = 
	\int {\cal D} A_0 \exp \left\{ i A^a_0 \left[ (D \cdot E)^a 
	- m_{\rm D}^2 W^a_{00}/\sqrt{4 \pi} \right] /T \right\} \, .
\eeq
Doing so makes $E$ and $W_{00}$ Gaussian as well, and they can now be
integrated out, yielding
\beqa
Z & = & \int {\cal D} A_i {\cal D} A_0 \exp ( - H' / T ) \, , \\
H' & = & \frac{1}{2} \int d^3 x \left[ B_i^a B_i^a + (D_iA_0)^a
	(D_iA_0)^a + m_{\rm D}^2 A_0^a A_0^a \right] \, ,
\eeqa
where the wave function term for $A_0$ arises from integrating out the
$E$ field and the Debye mass squared term arises from integrating out
$W_{00}$.  

The sole thermodynamic consequence of the $W$ fields is the
introduction of a Debye mass, and its magnitude is given exactly by the
coefficient in the $W$ field equations of motion.  This corresponds
exactly with what the complete hard thermal loop thermodynamic
contribution should be.  Furthermore, the Debye mass is introduced even
for $\lmax = 0$, the absolute minimum value.  We do not recommend using
$\lmax = 0$, however, because in this case the $W$ fields have no
dynamics and every $W^a_{00}$ is a conserved quantity.  Therefore, the
system is not ergodic and a Hamiltonian trajectory will not densely
sample the microcanonical ensemble.  However, to the best of our
knowledge the only conserved quantities (besides the Gauss constraints) 
in the non-abelian theory with
$\lmax \geq 1$ are energy and momentum\footnote{On a discrete lattice
the total momentum is not conserved, due to the Umklapp-effect.},
and we expect ergodicity in this
case.  
The conclusion is that the technique reproduces the thermodynamics of
the full HTL theory exactly, for all $\lmax \geq 1$.

\subsection{Propagator}

Now we turn to the study of the propagator in the $\lmax$ cut-off
theory.  We work only to linear order, or equivalently, we will
study the propagator only in the abelian theory.  In this case we can
study one $\vk$ mode in isolation.

Since the spherical harmonic expansion does not break rotational
invariance (even when we restrict $l\le \lmax$), it is sufficient to
study the propagation of modes for which $\vk$ is in the
3-direction.  The Fourier transformed equations of motion are
\ba
  \omega^2 A_m - k^2 A_m &=& \fr{m_{\rm D}^2}{3} W_{1m} \la{mom-A} \\
  \omega W_{lm} - k C_{lm,l'm',3} W_{l'm'} &=& 
	\omega \delta_{l,1}  A_m\,,  \la{mom-wlm}
\ea
where we have defined $A_{m=\pm 1} = \sqrt{4\pi/6}(\mp A_1 + iA_2)$ and
$A_{m=0} = \sqrt{4 \pi/3} A_3$.  The $A_{\pm 1}$ are the transverse
components of the gauge field and $A_{m=0}$ is longitudinal.

Since $C_{lm,l'm',3} \propto \delta_{m,m'}$, \cf~\eq\nr{clm-app}, the
equations of motion do not mix different $m$-sectors (this is the
advantage of choosing $\vk \parallel \hat e_3$).  We also note that
$W_{\lmax \lmax}$ and $W_{\lmax,-\lmax}$ do not evolve at all.
In general, the components with $m\ne \pm 1$ do
not couple to the transverse gauge fields.  We will not be concerned
here with the propagator in the longitudinal sector, or with any sector
which does not couple to any gauge fields, so
the only `interesting' modes are
those with $m=\pm 1$.  It should be noted that this decoupling
occurs {\em only\,} in the abelian theory.  (It also allows a more
efficient representation for hard thermal loops than the one we use
here, see \cite{Rajantie_HTL}.)

In the following we choose $m=1$, which is the sector which
couples to the transverse gauge fields.  The matrix $C_{ll'} = C_{l1,l'1,3}$
is a symmetric and traceless matrix of size $\lmax^2$ with non-zero
(positive) elements only if $l'=l\pm 1$.  (Note that, because $|m| \leq
l$, $l$ is restricted
here to the interval $1\le l \le \lmax$, hence the dimensionality of
$C_{ll'}$.)  As a result, in the eigenvalue problem
\be
  C \chi^{\alpha} = \lambda^{\alpha} \chi^{\alpha}\,,
\la{C-eigenvalue}
\ee
the eigenvalues $\lambda^{\alpha}$ are real and non-degenerate, 
and they come in positive and
negative pairs:  if $\lambda$ is an eigenvalue, so is $-\lambda$.
If $\lmax$ is odd, the matrix has one zero eigenvalue, otherwise
the eigenvalues are non-zero.  The eigenvectors $\chi^{\alpha}$ are real and
orthogonal, and we will normalize them to be orthonormal.

Writing the matrix $C_{ll'}$ in terms of the eigenvectors and eigenvalues,
\be
  C_{ll'} = \sum_{\alpha} \chi_l^{\alpha} 
	\lambda^{\alpha} \chi_{l'}^{\alpha}\, ,
\ee
we can solve for $W_{l1}$ in \eq\nr{mom-wlm}:
\be
  W_{l1} = \sum_{\alpha} 
	\fr {\omega}{\omega - k\lambda^{\alpha}} 
	\chi_l^{\alpha} \chi_1^{\alpha} A_1\,.
\ee
Inserting this in \eq\nr{mom-A}, we obtain the inverse transverse propagator
\be
 \Delta^{-1}_{\lmax} = -\omega^2 + k^2 + 
	\fr{m_{\rm D}^2}{3} \sum_{\alpha=1}^{\lmax} 
	\fr{\omega}{\omega-k\lambda^{\alpha}} (\chi_1^{\alpha})^2\, .
\la{invprop}
\ee

Let us now compare the propagator \nr{invprop} to the theory {\em
without\,} the $l$-cutoff.  Remember that in this case the propagator
has a {\em cut} in the interval $-k \le \omega \le k$
\cite{HTL}, and, in the limit $\omega \ll k\ll m_{\rm D}$, 
it describes {\em overdamped\,} behavior with damping coefficient
$\tau^{-1} \sim k^3/m_{\rm D}^2$.
Thus, the damping rate is $\sim g^4 T$ when $k \sim g^2T$, which is
the relevant momentum scale for non-perturbative physics.

What does the propagator look like at different values of $\lmax$?
If $\lmax=0$, the gauge fields are decoupled from the $W$ fields,
except through Gauss' law (see \eqs\nr{YM-equation} and
\nr{current-lm}); transverse physics is the same as in the absence of
the $W$ fields.
At $\lmax=1$ the ``matrix'' $C_{ll'}=0$ is a scalar, and
the propagator describes a massive vector particle:
$\Delta^{-1}_1 = -\omega^2 + k^2 + m_{\rm D}^2/3$.  

The first interesting case is $\lmax=2$.  The propagator is still easy
to solve analytically, and (using \eq\nr{clm-app}) the 
inverse propagator becomes
\be
  -\omega^2 + k^2 + \fr{m_{\rm D}^2}{3} \fr{\omega^2}{\omega^2 - k^2/5}\, .
\la{prop-l2}
\ee
The propagator has two zeroes given by $\omega^2 = k^2/5$, and
4 poles at
\be
  \omega^2 = \fr{3k^2}{5} + \fr{m_{\rm D}^2}{6} \pm \fr12
	\sqrt{\left(\fr{6k^2}{5} + \fr{m_{\rm D}^2}{3}\right)^2 - \fr{4k^4}{5}}\,.
\ee
In the limit $ k^2 \ll m_{\rm D}^2$ the poles are
\be
  \omega^2 = \fr{m_{\rm D}^2}{3} + \fr65 k^2 + O(k^4)\, ,   \h 
  \omega^2 = \fr35 \fr{k^4}{m_{\rm D}^2} + O(k^6)\,.
\la{lmax2}
\ee
The first 2 poles correspond to the plasmon, and give it the right
dispersion relation up to corrections
of order $k^4 / m_{\rm D}^2$.  The second 2 poles
are at $\omega \sim g^3T$ for $k\sim g^2T$ and $m_{\rm D}\sim gT$.  Thus,
instead of the correct overdamped behavior,
the $\lmax=2$ propagator \nr{prop-l2} describes {\em oscillatory\,}
behavior with $\omega \sim g^3 T$.  At first sight, this may
look like a fatal flaw in the $l$-mode cutoff method.  However, as we
will argue below, in practice this is not a serious drawback.

\begin{figure}[t]
\centerline{
\epsfysize=7.5cm\epsfbox{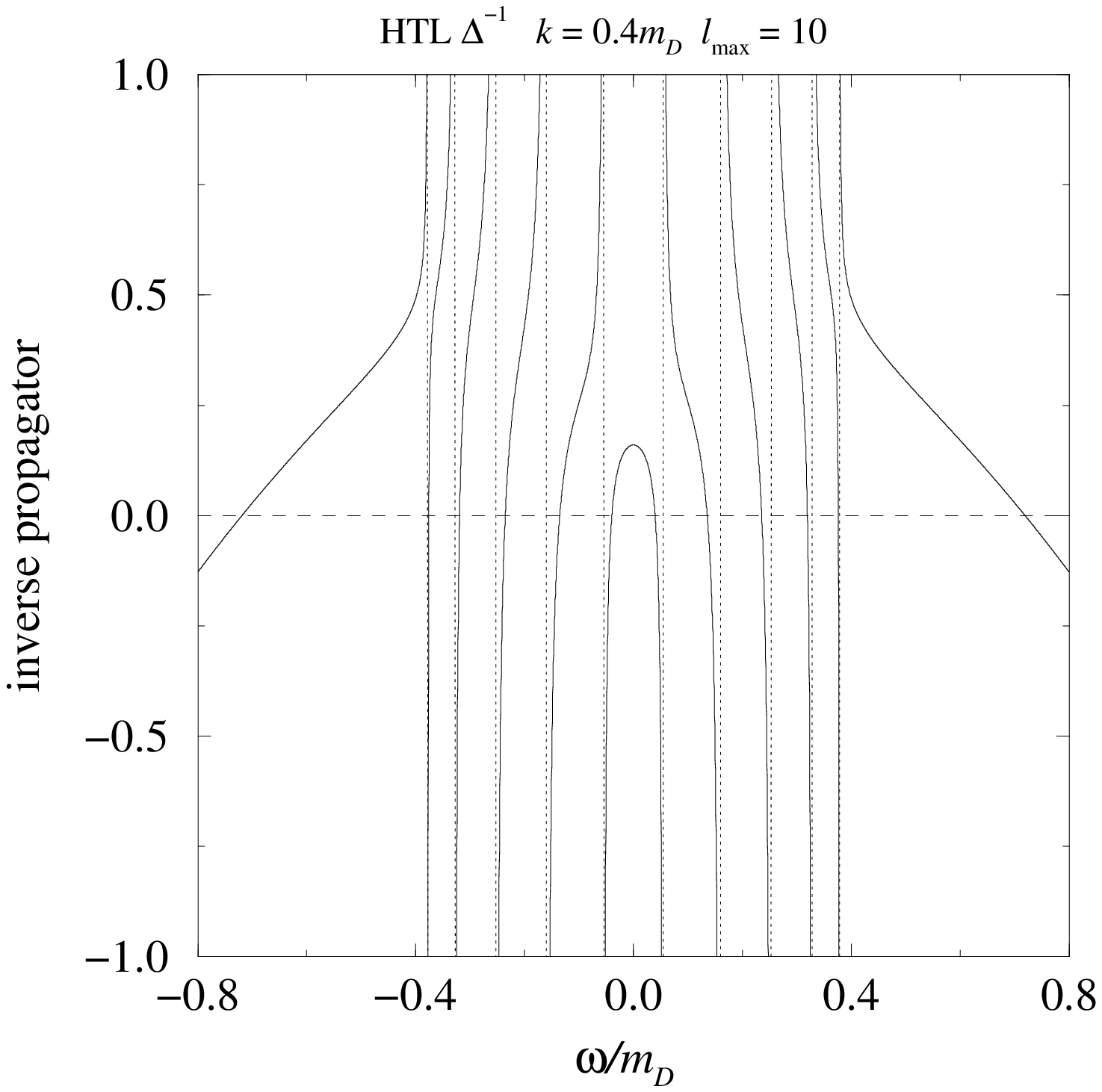}\hspace{4mm}
\epsfysize=7.5cm\epsfbox{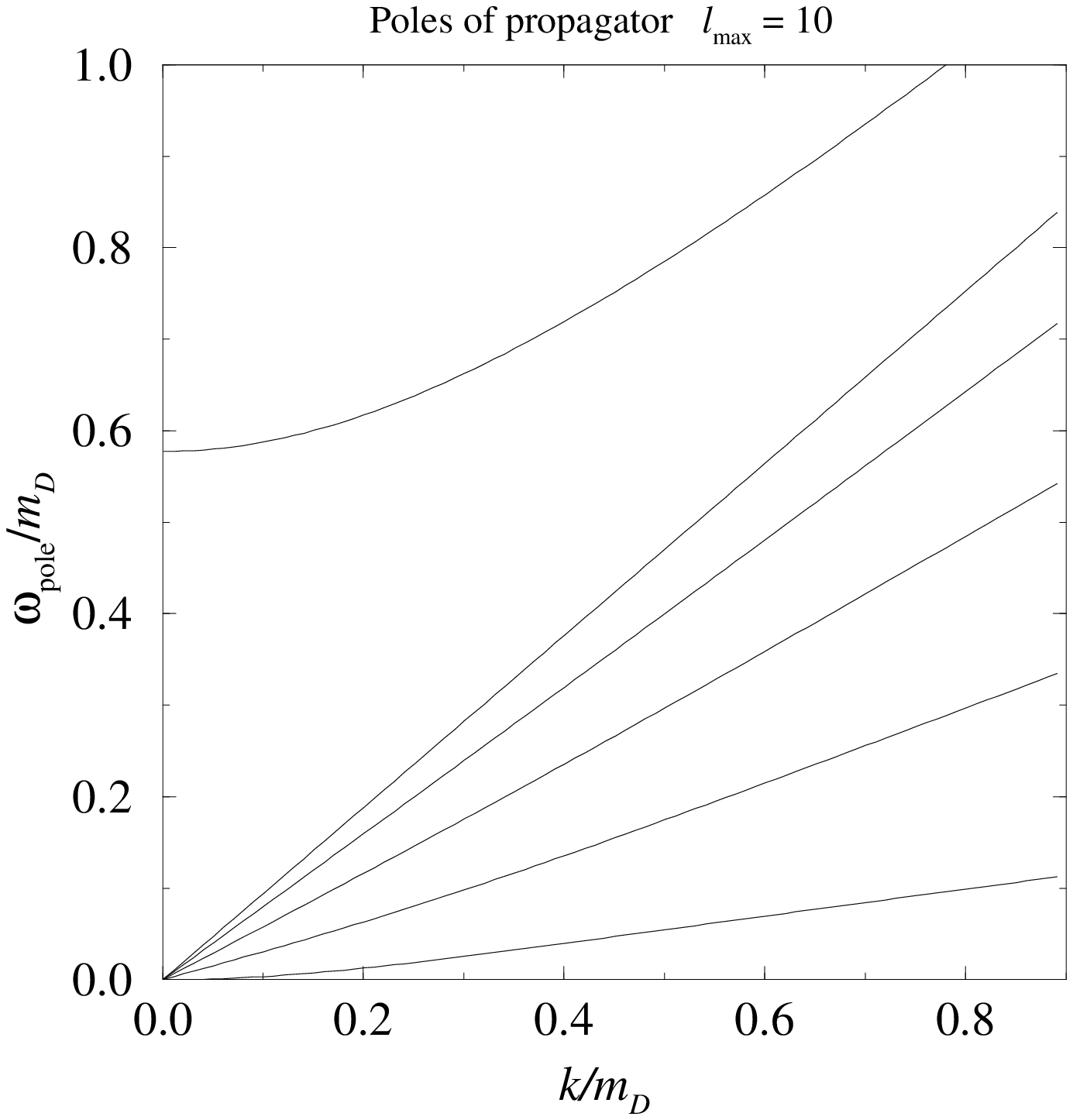}}
\caption[a]{{\em Left:}
The inverse propagator \eq\nr{invprop} with $\lmax=10$, plotted
against $\omega/m_{\rm D}$ with fixed $k=0.4 m_{\rm D}$.  {\em Right:} The
positive frequency poles of the propagator at $\lmax = 10$.  In these
figures, one can clearly see the development of the cut in the
interval $-k \le \omega \le k$, and the two plasmon poles at $\omega^2
\approx m^2_D/3 + 6k^2/5$.}
\la{fig:prop-l10}
\end{figure}

For odd values of $\lmax$ 
the matrix $C_{ll'}$ has one eigenvalue equal to zero.  As with $\lmax=1$, the 
self-energy contribution to \eq\nr{invprop} has a constant `mass term',
\be
	\fr{m_{\rm D}^2}{3}(\chi_1^{(0)})^2
	+\fr{m_{\rm D}^2}{3} \sum_{\alpha:\,\lambda^{\alpha}\ne 0}
	\fr{\omega}{\omega-k\lambda^{\alpha}} (\chi_1^{\alpha})^2\, .
\la{l-odd}
\ee
What this means is that there is a linear combination of $W$ and $A$
fields, namely $W_{l1}=W \chi^{(0)}_l, \; A = (m_{\rm D}^2/3k^2) W
\chi^{(0)}_1,$ which is strictly static.  Thus, part of the ``power'' in
the $A$ fields is lost to the dynamics of the system.  There are also
propagating modes, both at the plasmon frequency and for $\omega < k$.
For $\lmax=3$, the poles are at
\be
  \omega^2 = \fr{m_{\rm D}^2}{3} + \fr65 k^2 + O(k^4) \, , \h
  \omega^2 = \fr8{35} k^2 + O(k^4)\,.
\la{lmax3}
\ee
We can identify the same plasmon pole as with $\lmax=2$,
\eq\nr{lmax2}, but the other pole behaves as $|\omega| \sim k$ instead
of $|\omega|\sim k^2$.  For relevant values of $k$, the poles of
the $\lmax=2$ propagator are at much smaller $|\omega|$ than for
$\lmax=3$.

This pattern is seen to be true also for larger $\lmax$.
While we have been unable to find a general analytic expression for the
poles of the propagator, it is easy enough to solve the 
eigenvalue problem
\nr{C-eigenvalue} and find the poles of the propagators numerically.
In \fig\ref{fig:prop-l10} we show the inverse propagator and
the location of the poles when $\lmax=10$.
In general, we can state the following about the poles
of the propagator:
\begin{itemize}

\item[(i)] 
For $\lmax$ even, there are $\lmax$ poles and $\lmax$ zeroes of the
propagator in the interval $-k < \omega < k$.  For odd values of
$\lmax$, the number of poles and zeroes is $\lmax-1$.  In either case,
as $\lmax\rightarrow\infty$, the poles and zeroes merge into a cut in
the propagator.

\item[(ii)] 
There is a pair of plasmon poles at $\omega^2 \approx m_{\rm D}^2/3 + (6/5)
k^2 + O(k^4/m_{\rm D}^2)$.

\item[(iii)]
When $\lmax$ is even and $k \ll m_{\rm D}/\sqrt{\lmax}$, the lowest pair 
of poles behaves as 
\be
  \omega \approx \pm \fr{k^2}{m_{\rm D}\sqrt{\lmax}}\,,
\ee
whereas the other poles in the region $|\omega|<k$ depend linearly
on $k$.  For $\lmax$ odd, all of the poles in this region are linear.
As we make $\lmax$ larger, the power lost to the static mode becomes
smaller roughly as $\lmax^2$.
\end{itemize}

The absence of cuts
means that the gauge field propagation is non-dissipative.  We should
expect this behavior in the abelian theory because the equations are
linear.  However it need not concern us, because at large $\lmax$ the
behavior differs from the $\lmax=\infty$ limit only over very long time
scales, and the nonlinearities in the non-abelian case should become
important on shorter time scales if $\lmax$ is sufficiently large.

\begin{figure}[t]
\centerline{\epsfysize=9cm\epsfbox{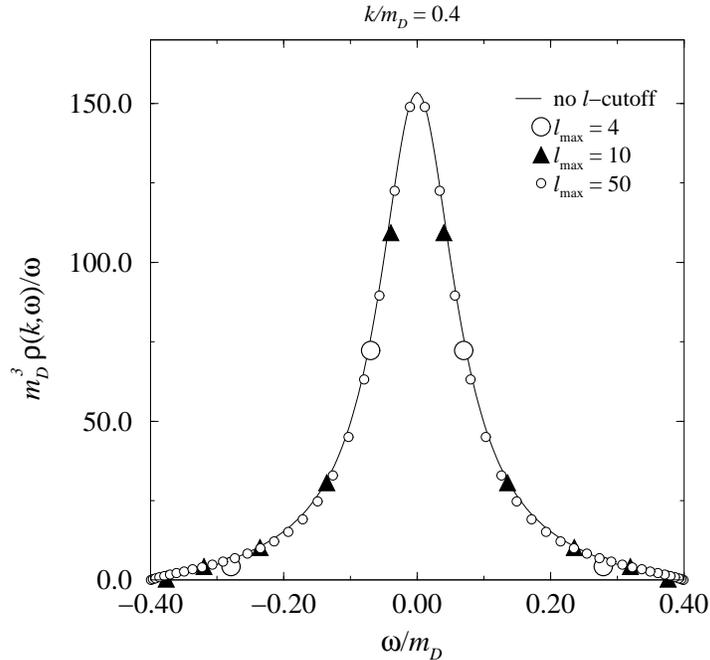}}
\vspace*{-5mm}
\caption[a]{
The spectral density $\rho/\omega$ at $k=0.4 m_{\rm D}$ for propagators
without the $l$-cutoff and with various values of $\lmax$.  The
spectral density for finite $\lmax$ is a sum of form $\sum_{\alpha} 
C_\alpha
\delta(\omega-\omega_\alpha)$.  The plot
symbols are plotted at coordinates $(\omega_\alpha, 
2C_\alpha/(\omega_{\alpha+1}-\omega_{\alpha-1}))$, 
which makes it possible to compare
the different $\lmax$-values.}
\la{fig:powersp}
\end{figure}
The spectral power density 
\be
\rho(\omega,\vk)/\omega = (2/\omega) \mbox{\,Im\,} \Delta 
(\omega+i\epsilon,\vk)
\ee
for fixed $k=0.4 m_{\rm D}$ is plotted in \fig\ref{fig:powersp}, both for the full
propagator without the $l$-cutoff and for several (even) values of $\lmax$.
In the finite $\lmax$ case the spectral density gets contributions only from
the poles of the propagator \nr{invprop}:
\be
  \rho_{\lmax}(\omega,k)/\omega = 
  \sum_{\rm poles} -\fr{2\pi}{\omega} 
	\delta(\omega-\omega_{\rm pole}(k))\times \mbox{Res\,}
\Delta_{\lmax}(\omega_{\rm pole}(k),k)\,.
\ee
The spectral power is strongly concentrated around $\omega = 0$ with a
peak width $\delta\omega \approx 4k^3/(\pi m_{\rm D}^2)$.  The spectral
power of the $\lmax$ propagator closely follows the $\lmax=\infty$
curve; however, in order to have enough power in the central peak
region, $\lmax$ should be large enough so that there are poles well
within the bulk of the peak, which is the relevant region for
the propagator to describe the correct damping.


We can use this property to derive an approximate `rule-of-thumb',
which tells how large $\lmax$ should be for a given value of
$k$ (which, for the relevant physics, should be set to $g^2T$):
we simply require that $\lmax$ is large enough so that lowest
positive frequency pole satisfies 
\be
  \omega_{\rm pole}(k) < \fr{4k^3}{\pi m_{\rm D}^2}\,.
\la{lmaxcrit}
\ee
Numerically, this corresponds to 
\be
  \lmax^{\rm even} > 0.62 m_{\rm D}^2/k^2 - 0.8, \h\h
  \lmax^{\rm odd}  > 1.86 m_{\rm D}^2/k^2 - 1.1 \,,
\la{required}
\ee
with good accuracy.
Strikingly, one has to use 3 times larger values for $\lmax$ in the
odd sector than in the even one.  This is due to the lack of the
$\omega \sim k^2$ -pole in the odd sector, as emphasized above.
While for modest values of $g^2T/m_{\rm D} = 0.3\ldots 0.5$, $\lmax = 6$ or
4 should be sufficient, for very weak coupling or large $m_{\rm D}$ the
required $\lmax$-value becomes impractical for numerical
work, since the numerical effort will rise as $(\lmax + 1)^2$.

Naturally, one has to remember that \eq\nr{required} is based on an
ad hoc requirement that the lowest positive frequency pole should be
within the peak of the spectral density, and different criteria would
lead to very different requirements (but the overall pattern in
\eq\nr{required} should remain).  What value of $\lmax$ one really
needs in non-abelian simulations may differ from
\eq\nr{required} by a large factor; indeed, the numerical results for
non-abelian theory in Sect.~\ref{sec:results} seem to imply that
\eq\nr{required} is overly strict.

We also note that the poles of the $W$ field coincide with the $A$ field
poles, so we have gotten them for free.
For odd $\lmax$, there is one pole not accounted for yet:  the mode
corresponding to the $C_{ll'}$ zero eigenvalue, which does not propagate
at all.  Naturally, the spectral power of the $W$ propagator is very
different from the $A$ propagator.

So, what do these results tell us about the sphaleron rate in the
non-abelian theory?  We conclude the following:

{\bf A.} When $\lmax$ is odd, a component of the gauge field is static
and not fluctuating, and therefore does not contribute to real time
processes.  Since the static component is largest in the infrared, we
expect this to reduce $\Gamma$ relative to
large $\lmax$ limit.  This behavior is worse for small
$\lmax$ and should go away at large $\lmax$ as the static component
contains less and less of the total gauge field amplitude.

{\bf B.}  In the even $\lmax$ sector, the
location and density of the poles
is relevant for the correct {\em damping\,} in hot plasma:  the larger
$\lmax$ is, the more the poles are able to reproduce the concentration
of spectral power at small $\omega$, and so the stronger the damping and
the smaller the sphaleron 
rate.  Thus, when $\lmax$ increases, the sphaleron rate should
approach the physical one from above.  The approach should be much
faster than in the odd $\lmax$ sector, see Eq. (\ref{required}).

This behavior is indeed close to what we observe in Sect.~\ref{sec:results}.

Obviously,  the results in this section imply that for fixed $\lmax$ one {\em
cannot\,} have the correct leading order (in $g$) behavior of the gauge field
propagator in the strict small $g$ limit.  
We expect this to be true also for the non-abelian gauge
propagator, and hence for the sphaleron rate.  Nevertheless, for
realistic values of $g$ and $m_{\rm D}$ we expect a modest $\lmax$ to be
sufficient.

\section{Lattice equations of motion}\la{sec:lattice}

In this section we discuss the discretization of the continuum
equations of motion \eqs\nr{YM-equation}, \nr{update-lm} and
\nr{current-lm}.  
Naturally, not all of the properties of the continuum evolution can be
satisfied on a discrete lattice, but the update rule of the lattice
system should fulfill at least the following criteria:

(i) Gauge invariance, lattice translational and rotational symmetry
and C, P, and T symmetries are preserved,

(ii) Gauss' law is identically satisfied,

(iii) The total energy is conserved.

Naturally, we also require that the small lattice spacing and smooth
field limit gives the correct continuum behavior.  

The discretization of the system is very similar to the pure Yang-Mills theory,
developed by Kogut and Susskind \cite{Kogut}.  The lattice is a
3-dimensional torus of size $L^3 = N^3\,a^3$, with lattice spacing
$a$.  As is customary in real-time simulations, we use $A_0 = 0$ 
gauge\footnote{We emphasize that this choice is just a convenient way to
fix the gauge ambiguity in the field update laws, and that any
alternative choice would give the same value for gauge invariant
correlators.} 
and discretize the gauge fields in terms of spatial parallel
transporters $U_i(x) = \exp(i g a A_i) \in \su2$, and electric fields
$E_i(x)$ which belong to the Lie algebra of \su2.  $U_i(x)$ and
$E_i(x)$ live on the links connecting points $x$ and $x+i$
(here we use the shorthand $x+i$ for $x + a\hat e_i$).
The $W_{lm}$ fields are located on lattice sites.  Thus, for each
lattice site the total number of field variables is 3 \su2-matrices and
$3+(\lmax+1)^2$ adjoint matrices.

On the lattice we want to use dimensionless field variables.
We absorb the lattice spacing and $g$ in lattice fields
as follows:
\be
    g a A   \rightarrow A\,, \h\h 
    g a^2 E \rightarrow E\,, \h\h
    g a W   \rightarrow W\, . 
\label{latt_scaling}
\ee
For compactness, we also use dimensionless lattice coordinates,
$x_i\rightarrow x_i a$, $x_i$ integer, reintroducing $a$ when
necessary.  We shall consider the evolution of the lattice fields both
in continuous and discrete time.  In discrete time, one update step
consists of evolving the fields from time $t$ to $t + \dt$, where $\dt
\ll 1$ in order to keep the evolution stable and integration errors
small.

\subsection{Gauge field update}

We shall use the standard single plaquette
definition for the magnetic field strength:
\be
  \fr1T \int d^3 x \fr14 F_{ij}^a F_{ij}^a 
  \rightarrow 
  \betaL \sum_{\Box} \big[ 1 - \half\tr U_\Box \big]\,.
\la{plaqb2}
\ee
Here $U_{\Box}$ is the ordered product of the link variables around
a plaquette, 
\be
  U_{\Box,ij}(x) = U_i(x) U_j(x+i) 
		  U^\dagger_i(x+j) U^\dagger_j(x)\,,
\la{plaquette}
\ee
At tree level $\betaL = 4/(g^2 T a)$.  However, this receives
radiative corrections; these will be discussed 
below.

Let us now consider the lattice gauge field equations of motion both
in continuous and discrete time.  The (continuous) time derivative of
the link matrix $U$ is given in terms of the electric field 
as\footnote{$E_i(x)$ appears on the left in Eq. (\protect{\ref{dU}})
because we choose to record $E_i(x)$ so that it transforms under
gauge fields as an adjoint object at the basepoint $x$ rather than the
endpoint $x+i$ of the link from $x$ to $x+i$.  Alternately we could work
in terms of $\tilde{E}_i(x) = U_i^{\dagger}(x) E_i(x) U_i(x)$, in which
case the expression would involve $U\tilde{E}$ rather than $EU$; similar changes
would appear in other expressions involving $E$.  There is no physical
difference between the two choices.}
\be
  \partial_t U_i(x,t) = i E_i(x,t) U_i(x,t)\,.
\la{dU}
\ee
The `gauge force' term in the evolution equation of the electric
field is fixed by the magnetic Hamiltonian \nr{plaqb2}, by varying
\eq\nr{plaqb2} with respect to  $A_i$.  When we add the current term due to 
the $W_{lm}$ fields, we obtain the evolution equation for
$E_i$:
\be
  \partial_t E^a_i(x,t) =
	- i\half\tr \bigg[ \tau^a U_i(x,t) 
		\sum_{|j|\ne i} S^\dagger_{ij}(x,t)\bigg] + 
	\fr12 \big[ j^a_i(x,t) + \trans^{ab}_i j^b_i(x+i,t) \big] \,. 
\la{dE}
\ee
Here $S_{ij}$ is the gauge link `staple' which connects the points
$x$ and $x+i$ around the plaquette:
\be
  S_{ij} = U_j(x) U_i(x+j) U^\dagger_j(x+i)
\la{staple}
\ee
The summation index $j$ in \eq\nr{estep} goes over both positive
and negative directions; a negative value means that
the link is traversed in the opposite direction as in \eq\nr{staple}:
$U_{-j}(x) = U_j^\dagger(x-j)$.

The current terms in \eq\nr{dE} are given by $j^a_i =
(m_{\rm D}a)^2/(4\pi) v^*_{mi} W^a_{1m}$.  Since $E_i(x)$ is located 
between the points $x$ and $x+i$, the current $j_i(x)$ is
averaged between the beginning and the end of the link.  The current
at $x+i$ has to be parallel transported to point $x$, and we use the
shorthand expression
\be
  \trans^{ab}_i\Phi^b(x+i,t)  = [U_i(x,t) \Phi(x+i,t) U^\dagger_i(x,t)]^a
\label{parallel-transp}
\ee
for the adjoint field parallel transport from point $x+i$ to point
$x$.\footnote{If we worked in terms of $\tilde{E}$, the other current
would require parallel transportation to the end point of the link.}

In discrete time, the adjoint field $E_i$ transports the link
matrix $U_i(t)$ to $U_i(t+\dt)$.  In order to keep the
evolution symmetric in time, it is natural to place $E_i$
in the half-timestep value $t+\half \dt$.  Integrating \eq\nr{dU},
we obtain the discrete time evolution equation for $U_i$:
\be
  U_i(x,t+\dt) = \exp\big[i E_i(x,t+\half\dt)\, \dt \big]\,  U_i(x,t)\,.
\la{ustep}
\ee
Alternatively, one can think of $\exp (i E_i\dt)$ as being the timelike
plaquette in the $(t,i)$ plane, which updates $U$ as shown because we
have chosen $A_0=0$ gauge.  (In another gauge there would be an extra
$A_0$ dependent term in the $U$ field update, and in the updates of the
$E$ and $W$ fields as well;
it is the convenience of leaving these out which encourages the choice
of temporal gauge.)

The discrete time electric field update can be obtained now
from \eq\nr{dE} by substituting
\be
  \partial_t E^a_i(x,t) \rightarrow 
	\fr1{\dt} [ E^a_i(x,t+\half\dt) - E^a_i(x,t-\half\dt) ]\,.
\la{estep}
\ee
The lhs of \eq\nr{dE} remains as is even at discrete time.  As formulated,
the discrete time update steps \nr{ustep} and \nr{estep} are symmetric
under time reversal and they give an algorithm 
accurate to order $O(\dt^2)$.  

As mentioned above, the relation $\betaL = 4/(g^2 T a)$ receives
corrections because UV modes behave differently on the lattice than in
the continuum.  This has been calculated in Appendix B of
\cite{particles} (see also \cite{Oapaper}), with the result
\ba
\betaL &=& \fr{4}{g^2 aT} + \left( \frac{1}{3} 
	+ \frac{37 \xi}{6 \pi}\right)  
        - \left(\frac{4}{3} + \frac{2 m^2_Da^2}{3}
	+ \frac{m^4_Da^4}{18} \right) \frac{\xi(m_{\rm D}a)}{4 \pi} \nonumber\\
&& {}	+ \left( \frac{1}{3} + \frac{m^2_Da^2}{18} \right) 
	\frac{\Sigma(m_{\rm D}a)}{4 \pi}\,.
\la{betaL}
\ea
Here $\xi = 0.152859\ldots$, and $\Sigma(m_{\rm D}a)$ and $\xi(m_{\rm D}a)$ are
integral functions:
\be
  \Sigma(m) = \int_{-\pi}^{\pi} \fr{d^3k}{(2\pi)^3}\fr{1}{\hat k^2 + m^2} \,,
  \h\h\h
  \xi(m) = \int_{-\pi}^{\pi} \fr{d^3k}{(2\pi)^3} \fr{1}{(\hat k^2 + m^2)^2}\,, 
\ee
where $\hat k^2 = \sum_i 4\sin^2 k_i/2$.  To 5\% accuracy,
this can be expressed as $\betaL \simeq 4/(g^2 aT) + 0.61$ for values
of $m_{\rm D}a$ used in this work.  However, we shall use the full
expression in our analysis.  In the sequel we will write $\beta_L$ for
the variable appearing in Eq. (\ref{betaL}), and write $\beta$ for $4 /
g^2 a T$.

Further subtleties related to this thermodynamic correction arise when
we convert $\Gamma$ to continuum limits; we will address this in
Appendix \ref{Oamatch}.

\subsection{$W_{lm}$ update and doublers}\la{sec:wlmupdate}

The $W_{lm}$ equation of motion \eq\nr{update-lm} has only first
order derivatives in time and space.  In order to preserve the 
exact $P$ and $T$ symmetries on the lattice, the first order derivative
terms should be replaced by {\em symmetric\,} finite differences.  Thus,
the continuous time lattice equation of motion for $W_{lm}$ is
\ba
 \partial_t W_{lm}(x,t) &=&   
  - \fr12\,C_{lm,l'm',i} [
      \trans_i W_{l'm'}(x+i,t) - \trans_{-i} W_{l'm'}(x-i,t)]  \nonumber \\
 & + &  \fr12 \delta_{l,1} v_{mi} [ E_{i}(x,t) + \trans_{-i} E_i(x-i,t) ] \,.
\la{dW}
\ea
The electric field contribution is symmetrized from each of the links
which connect to point $x$.  

As was done with the spatial derivative, we substitute the time
derivative $\partial_t W$ with a symmetric finite difference
$[W(t+\dt) - W(t-\dt)]/(2\dt)$, and the value of $W$ at time $t+\dt$
will depend on values at times $t$ and $t-\dt$. Explicitely, the
update rule becomes a `leapfrog' 
\ba
  W_{lm}(x,t+\dt) &=& W_{lm}(x,t-\dt) + 
   \dt \bigg\{ 2\delta_{l,1} v_{mi} E_{{\rm ave},i} \nonumber \\ 
    & - &  C_{lm,l'm',i} [
    \trans_i W_{l'm'}(x+i,t) - \trans_{-i} W_{l'm'}(x-i,t)] \bigg\} \,.
\la{wstep}
\ea
Here $E_{\rm ave}$ is the average electric field influencing the
propagation of $W_{lm}$ from $t-\dt$ to $t+\dt$.  Since this is over
two timesteps, there are 4 timelike `plaquettes' to each direction $i$:
\ba
  E_{{\rm ave},i}(x,t) &=& 
    \fr14 [ E_i(x,t-\half\dt) + \trans_{-i} E_i(x-i,t-\half\dt)  \nonumber \\
   & & ~{} +  E_i(x,t+\half\dt) + \trans_{-i} E_i(x-i,t+\half\dt)]\,.
\label{Eave}
\ea
Note that, due to \eq\nr{ustep}, the parallel transport $\trans_{-i}
E_i(x,t+\half\dt)$ can be made with $U$ matrices either at time $t$ or
time $t+\dt$ with the same result.  In practice, one does the $E_i$
transport once for each timestep, and stores the result for the next
timestep.  

To summarize, the discrete time update step ($t\rightarrow t+\dt$)
goes as follows:
\begin{itemize}
\item[1.] start with $U(t)$, $E(t - \dt/2)$, $W(t)$ and $W(t-\dt)$,
\item[2.] evaluate $E(t+\dt/2)$ with \eqs\nr{estep} and \nr{dE},
\item[3.] calculate $W(t+\dt)$ with \eq\nr{wstep} (and forget $W(t-\dt)$ and
	$E(t-\dt/2)$), and finally
\item[4.] calculate $U(t+\dt)$ with \eq\nr{ustep}.
\end{itemize}

A generic feature of a first order differential operator on a discrete
lattice is the decoupling of `odd' and `even' coordinate sectors:
$W_{lm}(x,t+\dt)$ depends only on $W_{lm}$ at points $(x,t-\dt)$ and
$(x\pm i,\dt)$; in particular it does {\em not\,} depend on
$W_{lm}(x,t)$, which is its immediate predecessor.  More precisely, if
we label the coordinates with an integer valued parity label $p =
\sum_i x_i + t/\dt$, the $W_{lm}$ fields at odd and even values of $p$
do not interact, except through their coupling to the gauge fields.
This causes a {\em species doubling problem}, in analogy to the one
familiar from lattice QCD (the Dirac equation is of first order).  The
properties of the doublers in a linearized theory are discussed in
detail in Appendix~\ref{app:doublers}.  

There are 15 extra low-energy doubler modes, living around the corners
of the 4-momentum space hypercube $k_i =(0,\pi/a)$, $\omega =
(0,\pi/(\dt a))$, with at least one of $k_i,\,\omega$ non-zero.  The
continuous time equation of motion $\nr{dW}$ has only 7 spatial
doublers; the rest are introduced by the time discretization (and can be
avoided, see subsection \ref{sec:nodoublers}).
However, in contrast to lattice QCD, in our case the doublers are
benign:  first, they couple only very weakly to the gauge fields,
decoupling completely at the corners of the Brillouin zone (see
Appendix~\ref{app:doublers}).  Second, they couple only
to gauge fields at very high wave numbers $k\sim 1/a$ and/or
frequencies $\omega \sim 1/(a \dt)$.  Thus, the doublers do not influence
at all the physically interesting small $k$ and $\omega$ gauge field
dynamics, and their effect on modes close to the lattice cutoff
remains small.  

Because the time step is small ($\dt \ll 1$), the timelike
doubler modes $\omega \sim \pi/a \dt$ are especially weakly coupled to
low-frequency modes.  Indeed, in simulations we used $\dt = 0.05$ and
observed no appreciable energy transfer between the timelike doublers
and low-frequency modes.  However, since the timelike doublers are
low-energy excitations of $W$ fields which are not present in
continuous time, they can cause problems in thermalization of the
system and, as it turns out, in counting the active degrees of
freedom.  This will be discussed below in
section~\ref{sec:thermalization}.

Before leaving the update we should comment on energy conservation.
In continuous time, we can write down the lattice version of the
Hamiltonian \nr{Hamiltonian-lm}:
\be
  H(t) = \betaL \sum_{\Box} \big[ 1 - \half\tr U_\Box(t) \big] 
   +  \fr12 \sum_{x,i} E_i^2(x,t) 
   +  \fr{(m_{\rm D} a)^2}{8\pi} \sum_{x,lm} |W_{lm}(x,t)|^2 \,.
\la{lattHt}
\ee
This Hamiltonian is exactly conserved by the equations of motion
\nr{dU}, \nr{dE} and \nr{dW}.  However, in discrete time there
is no equivalent conserved expression.  A good approximation
to the energy can be obtained by symmetrizing 
the contribution of the electric fields in \eq\nr{lattHt} with respect
to $t$:
\be
E_i^2(x,t) \rightarrow [E_i^2(x,t-\half\dt) + E_i^2(x,t+\half\dt)]/2 \,.
\ee
The energy obtained this way fluctuates with an amplitude $\propto
\dt^2$, but the mean value is stable.  The conservation of mean
energy is guaranteed by the time reversal symmetry of the discrete
time equations of motion:  if, at some point in the evolution of the
fields, we invert the sign of $E$ ($E \rightarrow -E$) and
conjugate and reverse sign for the hard particle charges
($W_{lm}\rightarrow -W^*_{lm} = -(-1)^lW_{lm}$), 
the system will exactly retrace
its evolution backwards.  If the energy had a tendency to increase,
inverting the time would cause it to decrease.  Since the
configurations $(U,E,W)$ and $(U,-E,-W^*)$ are just as likely to appear
in a thermal distribution, the system cannot exhibit any systematic
tendency for the average energy to change.  The stability of the
system is a necessary property for long Hamiltonian evolutions.

\subsection{Gauss' constraint}

The Gauss' law is given by the 0-component of the equations of the
motion \nr{YM-equation}:
\be
  D_i F^{i\,0} = j^0 = \fr{m_{\rm D}^2}{\sqrt{4\pi}} W_{00}\,.
\la{gausscont}
\ee
On the discrete spatial lattice and discrete time, care has to be
taken to make the appropriate symmetrizations to the fields $F_{i0} = E_i$
and $W$ appearing in \eq\nr{gausscont}.  Since $E_i$ is living
on half timestep time values $t + \half\dt$, we 
symmetrize $W_{00}$ from times $t$ and $t+\dt$:
\ba
 && \sum_i \bigg[ E_i(x,t+\half\dt) - 
	\trans_{-i}E_i(x-i,t+\half\dt) \bigg] \nonumber \\
 && \h {} + \fr{(m_{\rm D} a)^2}{\sqrt{4\pi}} \fr12 [ W_{00}(x,t) + W_{00}(x,t+\dt)] = 0
\la{gausslatt}
\ea

This condition (or rather, the constancy of the violation of this
condition) is satisfied exactly by the evolution equations
\nr{ustep}, \nr{estep} and \nr{wstep}.  To see this consider the change
of Eq. (\ref{gausslatt}) under one time step.  
It gets contributions from each
$dE/dt$ and from $dW_{00}/dt$.  (There are no contributions from the
time derivative $dU/dt$ of the $U$ appearing in the parallel transporter
$\trans_{-i}$ 
because $dU/dt$ commutes with $E$ and cancels between the $U$ and
$U^{\dagger}$ in Eq. (\ref{parallel-transp}).) 
In the absence of $W$ fields, the time derivative of Eq. (\ref{gausslatt}) 
is zero, as shown by Ambj{\o}rn and Krasnitz
\cite{AmbKras}.  The addition of $W$ fields adds new terms to the
$W_{00}$ field and $E$ field updates.
First there is a contribution to $E_i(x)$ and $\trans_{-i}E_i(x-i)$ from
$W_{1m}(x)$.  According to Eq. (\ref{dE}) these are equal; but $E_i(x)$
and $\trans{-i}E_i(x-i)$ appear in Eq. (\ref{gausslatt}) with opposite
sign, so there is no contribution here.  There is also no contribution
to $dW_{00}(x)/dt$ due to $W_{1m}(x)$.  Second, 
$W_{1m}$ at each
neighboring site contributes both to $dE/dt$ on the link between the
neighboring site and $x$, and to $dW_{00}/dt$, through
Eqs. (\ref{estep}) and (\ref{wstep}) respectively; but the two
contributions to the time derivative of Eq. (\ref{gausslatt})
cancel, because $C_{00,1m,i}=v^*_{mi}$.  Hence the update preserves
Gauss' law if it is satisfied by the initial conditions.  Enforcement of
Gauss' law is therefore a problem for the thermalization algorithm, not
the evolution.

\subsection{A way to eliminate temporal doublers}\la{sec:nodoublers}

There is an alternative way to write the update rules which eliminates
all the high frequency doubler modes, which we now discuss.  
First, note that the reason there are doublers is that the update as
specified in the previous subsections requires and maintains twice as
much information about the $W$ fields as is necessary.  As discussed in
the summary at the end of subsection \ref{sec:wlmupdate}, the update
needs the value of $W_{lm}$ at two time 
slices.
However, only $W$ and not its time derivative appear in the
Hamiltonian, so a complete specification of the fields should only
require $W_{lm}$ to be specified once at each site.  The excess
information describes the state of the doublers.  Eliminating the
doublers will require eliminating half of this information.
This is possible since, as noted
earlier, the update rule for $W$ does not mix the $W$ fields on odd and
even sublattices.  Therefore, it is possible to define $W$ only at every
other spacetime point; we can define it only at the even sites, that is, 
points for which $p=\left[ t/\dt + \sum_i x_i \right]$ 
is even.  Eq. (\ref{Eave}) remains unchanged, but
Eq. (\ref{estep}) has the modification
\beq
 \big[ j^a_i(x,t) + \trans^{ab}_i j^b_i(x+i,t) \big] \rightarrow
	\left\{ \begin{array}{cc} 
	j_i^a(x,t) \; , & \quad t/\dt + \sum_i x_i \; {\rm even} \\
	& \\
	{\cal P}_i^{ab} j_i^b(x+i,t) \; , & \quad
	t/\dt + \sum_i x_i \; {\rm odd} \\ \end{array} \right. \, ,
\eeq
that is, we use whichever $j$ is defined.  Similarly, in Gauss' law,
Eq. (\ref{gausslatt}) involves either $W_{00}(x,t)$ or
$W_{00}(x,t+\dt)$, whichever is defined.  The time derivative of
the Gauss constraint remains conserved, for the same reasons as before.

Updating the fields in this way removes 8 of the 15 doublers and cuts
the number of computations, and hence the CPU time, almost in half.  It
may slightly increase timestep errors because of the even-odd alternation
of the current in the $E$ field update rule; but this can be compensated
for by reducing $\dt$, which is not problematic because of the reduction
in the number of computations per time step.  We have compared the
update with and without this modification and find that the results for
physical measurables agree within statistical errors.

\subsection{Lattice thermodynamics} \la{sec:thermo}

In continuous time the equations of motion \nr{dU}, \nr{dE}, and
\nr{dW} describe a Hamiltonian evolution which conserves energy 
and phase space volume.  We can study the thermodynamics of the
system by using the Hamiltonian \nr{lattHt} to write down 
the canonical partition function
\be
Z = \int \bigg[ \prod_{x,i} dU_i(x) dE_i(x)\bigg]
	\bigg[ \prod_{x,lm} dW_{lm}(x)\bigg] 
     \prod_x \delta( G(x) )\, e^{-H/T}\,,
\ee
where $G(x)$ is Gauss' law, \eq\nr{gausslatt} (in continuous
time).  Introducing a Lagrange multiplier field $A_0$ in exact analogy
with what we did in continuous space in subsection \ref{thermo-subsec}, 
we can integrate out the $E$ and $W$ fields to obtain the lattice
partition function 
\ba
Z &=& \int \bigg[ \prod_{x,i} dU_i(x)\bigg] \bigg[\prod_x dA_0(x)\bigg]
        e^{-H_A}\,, \\
H_A &=& \betaL \sum_{\Box} \big[ 1 - \half\tr U_\Box(t) \big] + \nonumber \\
 &&	\fr12 \sum_{x,i} [\trans_i A^a_0(x+i)-A^a_0(x)]^2 + 
	\fr{(m_{\rm D}a)^2}2 \sum_x (A^a_0(x))^2 \,.
\la{HA}
\ea
The gradient term for the $A_0$ field is the simplest lattice
implementation of the continuum $(D_i A_0)^2$, and $m_{\rm D}^2$
appears as the $A_0$ mass term without any corrections,
just as in the continuum case.
The form of the partition function above is equivalent to the path
integral of the full quantum theory in the high-temperature
dimensional reduction approximation on the lattice \cite{KRS,KLRS}.
This guarantees that this theory reproduces the (equal time)
thermodynamics of the Yang-Mills fields.

This property can be used to fix the bare lattice value of the mass
term $m_{\rm D}$.  In general, classical field theories suffer from UV
divergences; however, when we consider the static thermodynamics of
the theory in \eq\nr{HA}, only a finite number of UV divergent
diagrams appears.  These divergences can be absorbed in counterterms,
and in particular for the theory in \eq\nr{HA}, we have \cite{KRS}
\be
  m^2_{\rm D,bare} = m^2_{\rm D,phys} -
	\fr{\Sigma g^2T}{\pi a},\h\h \Sigma = 3.17591\ldots\,.
\ee
Here $m^2_{\rm D, phys}$ is fixed according to the actual particle
content of the theory, see \eq\nr{mD}.  

\subsection{Thermalization}\la{sec:thermalization}

The real time simulation has to be started from a configuration which
has been chosen from a thermal distribution so that the the Gauss'
constraint is satisfied.  As emphasized above,
to start the update we need the fields $U(t)$, $E(t-\dt/2)$, $W(t)$
and $W(t-\dt)$.  

We will use the same general philosophy as in \cite{Moore1}.  Some of
the degrees of freedom, namely $E_i$ and $W_{lm}$, are Gaussian, while
others, namely $U_i$, are not.  We can draw the Gaussian fields from the
thermal ensemble and then use the evolution equations to ``mix'' this
thermalization with those degrees of freedom which are not Gaussian.
The thermalization proceeds by evolving the Hamiltonian equations of
motion of the system, but periodically
``refreshing'' the Gaussian degrees of freedom, that is, discarding
the values of Gaussian degrees of freedom and drawing them from the
thermal ensemble.

At first sight, this plan appears to be complicated due to the Gauss'
constraint.  In the case without the $W$ fields this problem was solved
in \cite{Moore1}, by first drawing $E$ from the Gaussian distribution
ignoring the constraint and then projecting to the constraint surface.
It is trivial to extend that technique to the current situation.
However it is actually possible to do something even easier.
Only the component
$W_{00}$ of $W_{lm}$ enters the Gauss' constraint.  Thus, according to
\eq\nr{lattHt}, we can set the higher $lm$-components freely to the
correct thermal distribution, that is, draw each of $W_{lm}^a$, $l\ge
1$, from a Gaussian distribution of width $\sqrt{8\pi/(m_{\rm D}a)^2}$.  The
thermalization then proceeds as follows:

\begin{itemize}
\item[(1)] Set $U(x,t)=1$, $E(x,t)=W_{00}(x,t)=0$. 

\item[(2)] Choose $W^a_{lm}(x,t)$, $l\ge 1$, from the Gaussian distribution
of width $\sqrt{8\pi/(m_{\rm D}a)^2}$.

\item[(3)] Evolve the equations of motion for a short period, transferring
energy from $W_{lm}$ to the other fields, while preserving Gauss'
law.

\item[(4)] Repeat from (2) until the fields are thermalized.
\end{itemize}

However, in discrete time we do not have an exact Hamiltonian, and
there is an inherent ambiguity $\propto \dt^2$ in the definition of
energy.  It is not immediately evident how the fields should be
thermalized.  At a more practical level, the randomization of $W_{lm}$
as above is complicated by the fact that we need $W_{lm}$ fields at
times $t$ and $t-\dt$ to start the leapfrog update.  
This is closely associated with the timelike doublers of the
$W$ fields.  However, as was discussed in section~\ref{sec:wlmupdate},
the timelike doublers couple extremely weakly to the low-frequency
mode sector, and there is practically no energy transfer between the
two sectors.  This was also seen in simulations:  the energy contained
in the doubler modes remained at the level where it was set by the
initial thermalization during the whole trajectory.\footnote{More
precisely, the energy transfer remains negligible for timestep
$\dt=0.05$ used in the simulations.  Using a dangerously large time step
of order $\dt\sim 0.2$--0.3, energy transfer becomes significant.}
Moreover, the gauge fields care only about the low frequency modes
(see Appendix~\ref{app:doublers}).  Thus, in principle, we are at
liberty to do whatever we choose about the timelike doubler modes; we
can either thermalize them or try not to excite them in thermalization.
The gauge fields will not see the difference --- however, in the
former case $W$ fields will contain roughly twice as much energy as in the
latter.  Note also that the whole problem would go away if we used the
update discussed in subsection \ref{sec:nodoublers}.

In all of our `production' runs we chose not to excite the timelike
doubler modes.  This makes the lattice modes resemble as closely as
possible the continuous time
fields.  Note that the Hamiltonian \nr{lattHt} counts the degrees
of freedom and energy equipartition correctly only if there are no
timelike doublers, and the $O(\dt^2)$ ambiguity in energy is valid
only in this case (in the presence of doublers, the ambiguity is of
order 100\%).

The thermalization without the doublers can be accomplished using the
steps (1)--(4) as above, but replacing the step (2), for example, by
one of the following two methods:

(a) Set $W_{lm}(t)$ to Gaussian random variables in step (2) above, and
perform the {\em first} update step in (3) using a forward asymmetric
time difference for these $lm$ modes:  that is, instead of
approximating the time derivative with $[W(t+\dt)-W(t-\dt)]/(2\dt)$ in
\eq\nr{wstep}, we use $[W(t+\dt)-W(t)]/\dt$.  This is a natural way to
start a leapfrog, and it gives a smooth interpolation for the fields.
This method gives slightly incorrect mean energy, but the error is
$O(\dt^2)$.

(b) Set $W_{lm}(x,t-\dt) = W_{lm}(x,t)$, where $l\ge 2$, to Gaussian
random variables in step (2).  Now also the first step can be
performed with the leapfrog.  This method excites the doublers more, but
the amplitude of their excitation is only $O(\dt^2)$.
Note that now only modes
$l\ge 2$ can be randomized, since in one timestep both $E$ and $W_{00}$
interact with with $l=0$ modes.

In our production runs we used the method (b).  We also made test runs with the
doubler modes fully exited.  This is simple to accomplish:  proceed as
in items (1)--(4) above, randomizing only $W_{lm}(t)$, $l\ge 2$, and
perform the evolution with the leapfrog update \nr{wstep}.  Since
there are now twice as many active $W_{lm}$ modes, the width of the
Gaussian distribution has to be multiplied by $\sqrt{2}$, in order for
the $U$ and $E$ fields to have the same total energy as before.  As
mentioned above, in the gauge field observables the doublers have no
observable effect.

Let us note that a Langevin-type thermalization, 
as used in \cite{Kras-therm}
for pure Yang-Mills theory, would be straightforward to implement
by coupling the noise to $W_{lm}$ fields.  Indeed, coupling the
noise only to the highest $l$-modes might be of interest even
during a simulation, since this could mimic the effect of the
higher $l$ modes.

\section{Measuring the Chern-Simons number diffusion}
\label{topology}

The baryon number violation rate is related to the diffusion
of the Chern-Simons number, defined as the charge associated
with the right-hand side of the anomaly equation \nr{fftilde}:
\be
 \ncs = \fr{g^2}{32\pi^2} \int d^3x \epsilon_{ijk}
 \left(F_{ij}^a A_k^a - \frac{1}{3} f_{abc} A^a_i A^b_j A^c_k \right) 
      = \fr1{N_G} \int d^3x J_B^0\,.
\ee
Since \su2 has a non-trivial third homotopy group $\pi_3(\su2) = {\bf
Z}$ the Chern-Simons number $\ncs$ is a topological index for vacuum
configurations:  we can perform any gauge transformation to a trivial
configuration $A=0$ without any cost in energy, and the resulting
configuration is as good a vacuum configuration as the initial one.
$\ncs$ is equal to the winding number of this gauge transformation.
Since it is now integer valued, it classifies the vacuum
configurations into disconnected classes, which cannot be continuously
gauge transformed to each other.  Thus, a vacuum-to-vacuum process
which increases $\ncs$ smoothly by one unit must go through a
non-vacuum exited state, the sphaleron.  Due to the axial coupling to
fermionic current, this process lifts one left-handed solution of the
Dirac operator from negative to positive energy, and pushes one right
handed state from positive to negative energy.  Since the \su2 sector
of the standard model is a chiral theory which does not couple to the
right-handed fermions, this process will create one fermion for each
fermionic generation ($N_G$).

At high temperatures the Chern-Simons number diffuses readily, and
integer values are not particularly preferred.  In any given
volume the Chern-Simons number performs a random walk in time, and the
diffusion constant, $\Gamma$, can be measured from
\be
  \Gamma = \lim_{V\rightarrow\infty}\lim_{t\rightarrow\infty}
	\fr{\langle (\ncs(t) - \ncs(0))^2\rangle}{Vt}\,.
\ee
Here the angle brackets $\langle\cdot\rangle$ refer to an average over
the thermal ensemble.  The change in the Chern-Simons number can be
evaluated from
\be
 \ncs(t)-\ncs(t_0) = \fr{g^2}{8\pi^2} \int_{t_0}^t dt' 
	\int d^3x  E_i^a B_i^a\,.
  \la{ncsint}
\ee
In principle, this measurement is readily convertible to lattice
language:  lattice versions of the fields $E$ and $B$ feature prominently
in the equations of motion \nr{dU},\nr{dE}.  However, this ``naive''
definition of $\ncs$ on the lattice, often used in the early work on
Chern-Simons number diffusion in lattice \su2 gauge theory
\cite{Ambjornetal,AmbKras,Moore1,TangSmit}, suffers from spurious
noise and diffusion which obscures the physical $\ncs$ diffusion.
Moreover, due to its UV nature, the amplitude of the noise diverges as
$1/a$ in fixed physical volume, which is disastrous in the continuum
limit.  The reason for this noise is well understood:  the integral
over lattice $\bf E\cdot B$ on the right-hand side of
\eq\nr{ncsint} does not form a total time derivative, and hence it
depends on the path along which one connects the initial and final
configurations in \eq\nr{ncsint}.  In other words, it does not give us
a topological measurement.

In general, topology of lattice fields is
ambiguous, since the variables are always continuously
connected to trivial ones.  However, at fine enough lattice spacings
(still easy to achieve in our simulations) almost every one of the
plaquettes is very close to unity in a thermal ensemble; large
plaquette values are exponentially suppressed.  Perturbatively this
means that the gauge fields are small, and for this subset of lattice
fields topology can be unambiguously defined \cite{Luscher}.  
Physically, this
means that the spatial size of the topology changing configurations,
sphalerons, is large in lattice units.  This will be true because the
energy of sphaleron-like configurations increases linearly with inverse size.
The Boltzmann suppression factor for small (lattice
scale) sphalerons is enormous and they, in practice, never appear in
simulations.  The interplay between entropy and the
Boltzmann factor sets the dominant sphaleron size to be $\sim g^2 T$.
For topology to be unambiguously defined, 
our lattice spacing must be considerably smaller than this.

Two successful methods for measuring topology in the current ``real
time'' context have been recently developed.  The first method uses
an auxiliary 
``slave field'' to track the winding number of the gauge
\cite{slavepaper}.  It is a development of a method originally
proposed by Woit \cite{Woit}, which is based on counting the winding
numbers of singularities in the Coulomb gauge.  In this work we use the
second method, ``calibrated cooling.''  This method is based on
the cooling method by Ambj{\o}rn and Krasnitz \cite{AmbKras2},
and fully developed by Moore in \cite{broken_nonpert}.  
The rest of this section will summarize this method.

The calibrated cooling method relies directly on the fact that the
sphalerons are large and extend over several lattice units.  Thus, we
can get rid of most of the ultraviolet noise in the thermal
configuration by applying a small amount of {\em cooling\,} to the
configuration:  the resulting configurations are very smooth on lattice
scales, but they still have the same topological content as the
original configuration.  After cooling 
the integral \nr{ncsint} can be performed
with small errors.  The accumulation of residual errors is
prevented by periodically cooling all the way to a vacuum
configuration:  we know that the true vacuum-to-vacuum $\delta\ncs =
\mbox{integer}$, and any deviation is due to accumulated
integration error, which can thus be ``calibrated'' away.  This
is schematically described in \fig\ref{fig:cooling}.

\begin{figure}[t]
\centerline{\epsfxsize=11cm\epsfbox{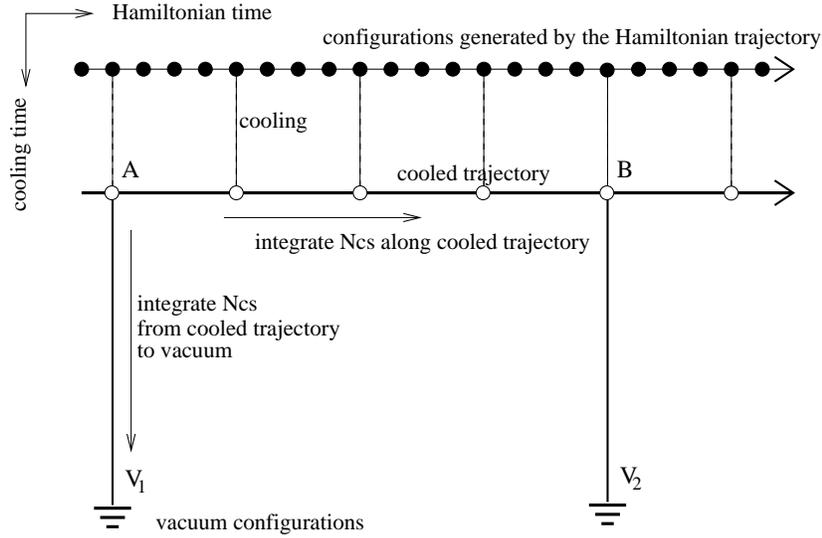}}
\caption[a]{
How the $\ncs$ evolution is measured (after \cite{broken_nonpert}).
Top horizontal line shows the configurations (solid circles) generated
by the lattice equations of motion.  Every few timesteps, the
configurations are cooled a fixed cooling length, giving a parallel
cooled trajectory (open circles).  Now the fields are smooth enough so
that ${\bf E\cdot B}$ can be reliably integrated, giving
$\delta\ncs(t)$ along the cooled trajectory.  In longer intervals, the
$\ncs$ measurement is ``grounded'' by cooling all the way to a vacuum
configuration.  If $\delta\ncs$ along paths like $V_1\rightarrow A
\rightarrow B \rightarrow V_2$ is always close to an integer, we know
that the integration errors are small.  The residual deviation from an
integer value is subtracted from $\delta\ncs(A\rightarrow B)$,
cancelling the accumulation of errors.\la{fig:cooling}}
\end{figure}

The cooling path is defined by the gradient flow of the standard
single plaquette Kogut-Susskind gauge action, given in \eq\nr{plaqb2}.
The evolution along this path is parametrized by fictitious cooling
``time'' $\tau$, dimensionally $\mbox{(length)}^2$.  The cooling
equation of motion is now \cite{AmbKras2}
\be
   \fr{\partial U_i(x)}{\partial \tau} 
   = i\sigma^a \fr{\partial A^a_i}{\partial\tau} U_i(x)
   = i\sigma^a \half\tr \bigg[ i\sigma^a U_i(x) 
		\sum_{|j|\ne i} S^\dagger_{ij}(x)\bigg] U_i(x)\,.
\la{cooling}
\ee
Here the staple $S$ is defined as in \eq\nr{dE}.  On the lattice the
equation above is evolved in discrete $\tau$, and we use here
optimized step lengths by alternating $\delta\tau/a^2 = 5/48$ and $10/48$.
Too large a time step causes the UV modes to become
unstable.  

The evolution of \nr{cooling} all the way to a vacuum configuration is
a computationally demanding task, and it can easily dominate the
cpu time.  The integration can be dramatically accelerated by {\em
blocking\,} the lattice:  after a bit of cooling the fields are very
smooth at the lattice scale, and essentially no information is lost if
we reduce the number of lattice points in each direction by a factor
of 2.  The blocked $U$-matrices are formed from the product of the
matrices on the two links between the blocked points.  Since the
lattice spacing $a$ is now increased by a factor of 2, computational
cost in cooling is reduced by a factor $2^5$ -- a factor of 4 coming
from the increase in the $\delta\tau$ step.  In our calculations we
block the configuration twice in the course of cooling to vacuum.
For detailed information about this method we refer to \cite{broken_nonpert}.

In all of our simulations we cooled a configuration from the
Hamiltonian trajectory at intervals $\delta t = 0.5 a$ (once in 10
timesteps with $\dt=0.05 a$).  We cooled to depth $\tau = a^2 45/48$
for the cooled trajectory (see \fig\ref{fig:cooling}), using unblocked
configurations.  $\delta\ncs$ was integrated along this trajectory
using improved $O(a^2)$ accurate definitions for ${\bf E\cdot B}$
\cite{Moore1}.  The cooling to vacuum was performed with an interval
$\delta t = 12.5$.  Our parameter choices were overly conservative:
the vacuum-to-vacuum integration error was typically of order
0.02--0.04.  Thus, it is possible to use much more aggressive
optimization than we use here without losing the topological nature
of this measurement (see ref.~\cite{MooreRummukainen}).

\section{Simulations and results} \la{sec:results}

Our aim here is to answer the following questions: 

\begin{itemize}

\item[(1)] what is the dependence of the Chern-Simons number diffusion rate
$\Gamma$ on the finite $\lmax$ cutoff, and is there an $\lmax$
which is `large enough' for practical purposes or is an
$\lmax\rightarrow\infty$ extrapolation necessary?

\item[(2)]
 is $\Gamma$, in physical units, independent of the lattice spacing?

\item[(3)] how does $\Gamma$ depend on the physical quantity $m_{\rm D}/g^2T$?

\end{itemize}
Let us first discuss the relation between the physical Debye mass
$m_{\rm D}$ and the bare mass parameter $m_{\rm D}a$ on the lattice.  As explained
in Sect.~\ref{sec:thermo}, the bare mass receives renormalization
counterterms and diverges in the UV limit as $1/a$.  However,
according to the scaling arguments of Arnold, Son and Yaffe
\cite{ASY}, the sphaleron rate should not actually depend on the Debye
mass, which characterizes static screening properties of the hot
plasma, but on the damping rate of the transverse gauge field
propagation.  As explained in Sect.~\ref{Propagator-sec}, this is
related to the Debye mass in the continuum.  However, due to the
lattice dispersion relation, the hard gauge field modes do not
propagate at the speed of light, and their effect on the damping is
reduced.  Averaging over all of the directions of the lattice momenta,
Arnold \cite{Arnoldlatt} has calculated that the effect of the hard gauge
field modes on the lattice is a factor of ($0.68\pm0.2$) times smaller
than the continuum relation between the damping coefficient and $m_{\rm D}^2$
would imply.  The error quoted is systematic, and it takes into
account the rotational non-invariance of the lattice propagators.
Thus, we shall use the following relation between the bare lattice
$m_{\rm D}$ and the continuum one:
\be
  Z^{-1}_{m_D} m_{D,\rm latt}^2 = 
	m_{D,\rm phys}^2 - 0.68 \fr{\Sigma g^2 T}{\pi a}\,.
\la{mdlatt}
\ee
Here $Z^{-1}_{m_D}$ is a radiative correction of form $1+O(a)$, see
\eq\nr{Zmd}.  We use the improved relation \eq\nr{betaL} to relate the
lattice spacing $a$ to the physical scale $g^2T$.  There are
additional radiative corrections associated with renormalization of
the lattice time scale, which we discuss in Appendix \ref{Oamatch}.
In order to avoid the uncertainties associated with the UV
counterterm, we use mostly fairly large physical values of $m_{\rm
D}$ so that the UV term remains subdominant.  The results are actually
quite robust against variations in the numerical coefficient 0.68\@,
even with the smallest $m_{\rm D}$ we use.

\begin{table}[t]
\centerline{
\begin{tabular}{|l|rr|l|} \hline
run parameters          & $\lmax$ & time$/a$ & $\Gamma/(\alpha^4 T^4)$    \\ 
\hline 
$\betaL = 8.7$, $V/a^3 = 24^3$       & 0 &  8000 & 1.49(15)    \\ 
$m^2_D = 1.59/a^2 = 8.20 g^4T^2$ ~~~  & 1 & 20000 & 0.0606(70)  \\
                                      & 2 & 20000 & 0.531(34)   \\
                                      & 3 & 30000 & 0.345(23)   \\
                                      & 4 & 20000 & 0.520(22)   \\
                                      & 5 & 20000 & 0.445(30)   \\
                                      & 6 & 30000 & 0.534(27)   \\
                                      &10 & 20000 & 0.518(34)   \\
\hline
$\betaL = 12.7$, $V/a^3 = 32^3$       & 2 & 37500 & 0.249(23)   \\
$m^2_D = 1.98/a^2 = 20.7 g^4T^2$      & 4 & 37500 & 0.198(18)   \\
                                      & 6 & 45000 & 0.199(18)   \\
\hline
$\betaL = 12.7$, $V/a^3 = 32^3$       & 2 & 37500 &  0.839(36) \\
$m^2_D = 0.29/a^2 = 4.84 g^4T^2 $     & 4 & 37500 &  0.687(50) \\
\hline
\end{tabular}}
\caption[0]{How the sphaleron rate $\Gamma$ depends on $\lmax$.}
\la{tab:lmax}
\end{table}

\paragraph{$\lmax$ dependence:}
In order to study how the sphaleron rate depends on the value of
$\lmax$, we performed a series of runs with $24^3$ lattices using
fixed $\betaL = 8.7$ and $m_{\rm D}^2 = 1.5/a^2 = 7.9 g^4T^2$, and varied
$\lmax$ from 0 to 10, as shown in Table~\ref{tab:lmax}.  The results
are also shown in \fig\ref{fig:lmax}.  When $\lmax$ is even, the
results are remarkably stable:  indeed, the data from $\lmax=2$ to 10
are mutually compatible within the statistical errors.
However, for odd $\lmax$ the rate remains substantially smaller,
approaching the even sector value from below when $\lmax$ increases.

\begin{figure}[t]
\centerline{\epsfysize=10cm\epsfbox{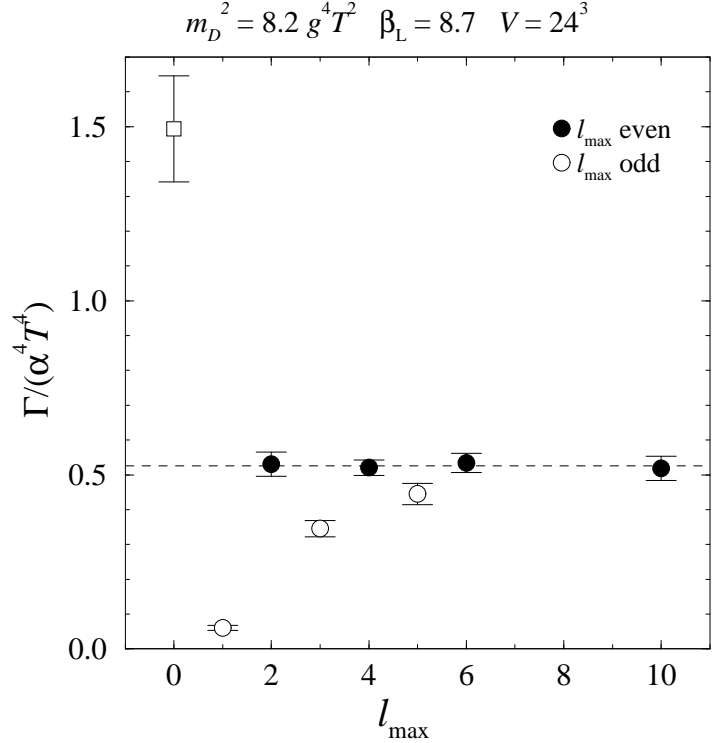}}
\caption[a]{The dependence of $\Gamma$ on $\lmax$ on a lattice of size
$24^3$,$\betaL = 8.7$.} 
\la{fig:lmax}
\end{figure}

The special case $\lmax = 0$ has a rate which is $\sim 3$ times larger
than the $\lmax=2,4,\ldots$ rate.  This is actually close to
the rate measured from standard \su2 gauge theory without any $W$
fields at the same $\betaL$ \cite{MooreRummukainen}; there the rate was
$1.68 \pm .03$.

This behavior is qualitatively in accord with the theoretical analysis
in the abelian theory in Sect.~\ref{Propagator-sec}.  The odd $\lmax$
sector gives a substantially reduced rate because much of the infrared
power is in non-propagating modes, and is therefore not available to
participate in Chern-Simons number diffusion.  However, for the even
$\lmax$ sector we do not see the gradual decrease in the rate as
predicted by the analysis in Sect.~\ref{Propagator-sec}, the rate just
snaps to the correct level immediately when the damping is turned on
by going from $\lmax=0$ to $\lmax=2$.  According to the requirement
for minimum $\lmax$ given in \eq\nr{required}, we should use $\lmax
\gsim 0.62 m_{\rm D}^2/g^4T^2 - 0.8 \approx 7$ (for even $\lmax$).
The naive limits given in \eq\nr{required} are obviously too strict
for the non-abelian theory.

At larger $m_{\rm D}^2/g^4T^2$ the difference between $\lmax = 2$ and higher
values should be more visible.  Indeed, in simulations at
$m_{\rm D}^2/g^4T^2 = 20.1$, using $\lmax = 2$, 4 and 6\@, we do observe a
significant decrease in the rate as $\lmax$ increases from 2 to 4;
this is shown in Table~\ref{tab:lmax}.  Here we use a smaller lattice
spacing, $\betaL = 12.7$, and correspondingly larger volume in lattice
units.  In this case the required $\lmax$, according to
\eq\nr{required}, would be $\sim 12$.
We also see an effect in the rate at $m_{\rm D}^2/g^4T^2 = 4.75$,
$\betaL=12.7$, using $\lmax=2$ and 4.

\paragraph{Physical sphaleron rate:}

\begin{figure}[t]
\centerline{\epsfysize=10cm\epsfbox{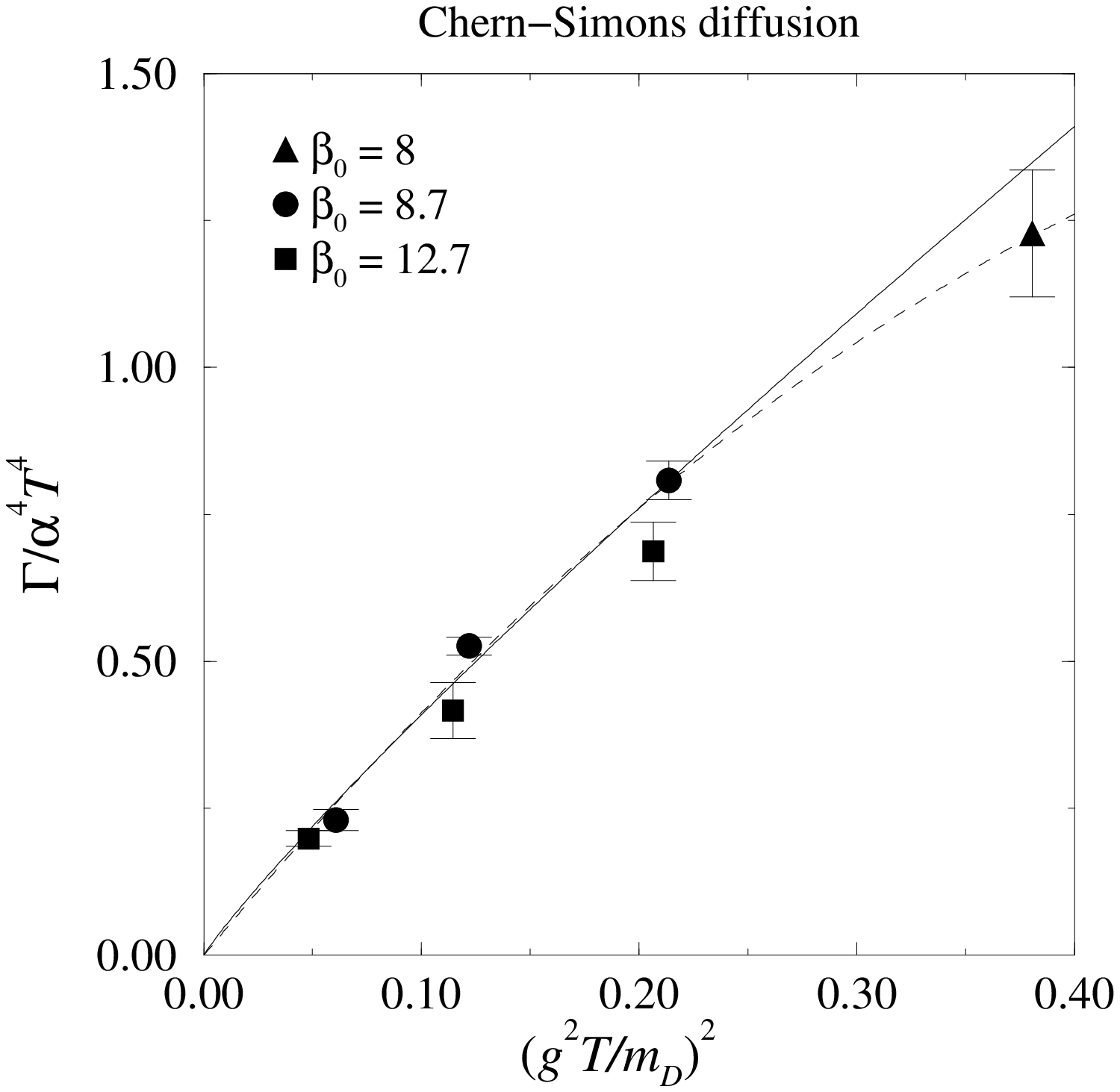}}
\caption[a]{The Chern-Simons number diffusion rate $\Gamma$ in physical
units.  Dashed line:  fit to linear + second order term, \eq\nr{fit1};
continuous line:  fit to linear + a log-term, \eq\nr{fit2}.}
\la{fig:kappa}
\end{figure}

\begin{table}[t]
\centerline{
\begin{tabular}{|rrc|ll|l|ll|} \hline
$\betaL$ & $\lmax$ & $V$ & $(m_{\rm D} a)^2$ & $ m_{\rm D}^2/g^4T^2$ & time$/a $ 
	& $\Gamma/\alpha^4T^4$ & $\kappa'$ \\ \hline 
8.0  &        4 &    $24^3$ &   0.375    &  2.63
& 10000 & 1.23(11)  & 40.5(3.6)\\
\hline
8.7  &        4 &    $24^3$ &   0.766    &  4.68  
& 37500 & 0.808(33) & 47.5(1.9)\\
8.7  & 2,4,6,10 &    $24^3$ &   1.59     &  8.20  
& 90000 & 0.526(15) & 54.2(1.6)\\ 
8.7  &        4 &    $24^3$ &   3.51     & 16.4  
& 25000 & 0.230(18) & 47.5(3.7)\\
\hline
12.7 &       4  &    $32^3$ &   0.291    &  4.84  
& 37500 & 0.687(31) & 46.8(1.9)\\
12.7 &       6  &    $32^3$ &   0.707    &  8.74  
& 20000 & 0.417(48) & 45.8(5.2)\\
12.7 &     4,6  &    $32^3$ &   1.97     & 20.7  
& 82500 & 0.199(13) & 51.7(3.5)\\
\hline
\end{tabular}}
\caption[0]{The Chern-Simons diffusion rate $\Gamma$ and the parameter
of the Arnold--Son--Yaffe scaling law $\kappa'$, \eq\nr{asy-eq}.  If
the results at different values of $\lmax$ are statistically compatible, 
we have taken an average over them, as indicated.
}\la{tab:results}
\end{table}

\begin{figure}[t]
\centerline{\epsfysize=10cm\epsfbox{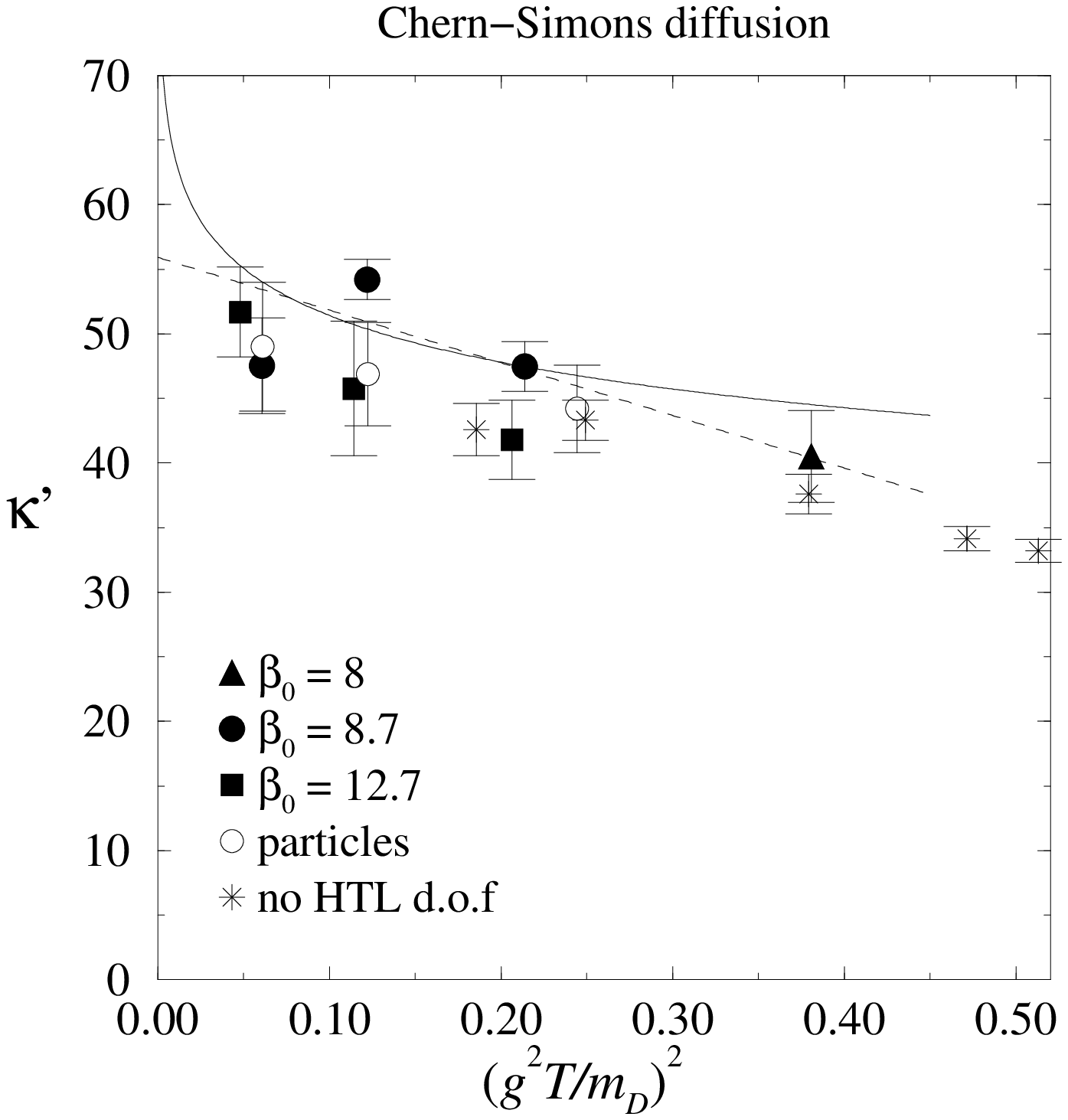}}
\caption[a]{The Chern-Simons number diffusion rate $\Gamma$, expressed
as $\kappa' = (\Gamma/\alpha^5T^4) (m_{\rm D}^2/(g^2T^2))$.  Dashed
line: fit to linear + second order term, \eq\nr{fit1}; continuous
line: fit to linear + a log-term, \eq\nr{fit2}.  For comparison, we
include the results obtained with the `particles' method
\cite{particles} and also without any hard thermal loop degrees of freedom
\cite{MooreRummukainen}.  These points are not included in the fits.}
\la{fig:kappaprime}
\end{figure}

The results for the sphaleron rate are shown in
Table~\ref{tab:results}, and plotted in Figs.~\ref{fig:kappa} and
\ref{fig:kappaprime}.  The ``old argument''
\cite{ArnoldMcLerran,KhlebnikovShaposhnikov}, 
based on dimensional analysis, says
that the rate should scale as $\Gamma = \kappa \alpha^4 T^4$ with
$\kappa$ a constant.  This behavior is clearly excluded, as already seen
in \cite{particles,MooreRummukainen}.  Rather, the rate falls linearly 
in $g^4T^2/m_{\rm D}^2$, confirming
the ASY scaling picture.  Indeed, we can make a fit of form
\be
   \fr{\Gamma}{\alpha^4T^4} = 
	c_1 \fr{g^4T^2}{m_{\rm D}^2} + c_2 \left(\fr{g^4T^2}{m_{\rm D}^2}\right)^2
\la{fit1}
\ee
to the data, with the result $c_1 = 4.5\pm0.2$, $c_2 = -3.2\pm 1.1$,
with $\chi^2 = 12$ for 5 degrees of freedom.  Within our statistical
errors, we did not observe any systematic lattice spacing dependence,
and we use the results obtained with all the lattices in
Table~\ref{tab:results} in the fit.  If we include the known
logarithmic contribution \cite{Bodek_paper},
\be
  \Gamma_{\log} = (0.425 \pm 0.027) \fr{g^4T^2}{m_{\rm D}^2} 
\log\left(\fr{m^2_D}{g^4T^2}\right) \alpha^4 T^4\,,
\ee
we can perform a one-parameter fit
\be
   \fr{\Gamma}{\alpha^4T^4} = 
        \fr{g^4T^2}{m_{\rm D}^2} 
	\left[ 0.425 \log \left(\fr{m^2_D}{g^4T^2}\right) + d \right]
\la{fit2}
\ee
with $d= 3.09 \pm 0.08$, with $\chi^2 = 15$ for 6 degrees of freedom.
The logarithmic contribution actually makes the fit a bit worse;
however, a subleading term $O(g^4T^2/m_{\rm D}^2)$ would not change
it, since its coefficient would be compatible with zero.  The errors
quoted above are purely statistical; we shall discuss systematic
errors below.

The rate becomes approximately constant, and the difference between
the two fits becomes more visible, if we plot the rate in terms of
the coefficient of the ASY scaling law $\kappa'$, defined through
\be
  \Gamma = \kappa' \fr{g^2 T^2}{m_{\rm D}^2} \alpha^5 T^4\,.
\la{asy-eq}
\ee
The values of $\kappa'$ are given in Table~\ref{tab:results} and shown
in \fig\ref{fig:kappaprime}.  We also include here the data calculated
by Moore, Hu and M\"uller with the particle degrees of freedom
inducing the hard thermal loop effects \cite{particles},
and the results obtained by Moore and Rummukainen using only SU(2)
gauge fields without any additional hard thermal loop degrees of
freedom \cite{MooreRummukainen}.  In the latter case the damping
arises solely through the UV gauge field modes on the lattice, and
$m_{\rm D}^2$ can be obtained from \eq\nr{mdlatt} by setting $m_{\rm D, latt}^2
= 0$.

The consistency of the results obtained with different methods in
\fig\ref{fig:kappaprime} is remarkable.  This gives strong credibility
to the view that the damping of the sphaleron rate seen in the
simulations is really caused by physical hard thermal loop effects.
Perhaps surprisingly, the pure Yang-Mills results are perfectly in
line (within the statistical errors) with the results obtained with
the hard thermal loop effective theories, even though in the former
case the spectrum of the hard modes is strongly distorted by the
lattice dispersion relation
\cite{Arnoldlatt}.  Also, the consistent decrease of $\kappa'$ with
increasing $(g^2T/m_D)^2$ in
\fig\ref{fig:kappaprime} strongly suggests that this
subleading effect is not due to lattice effects,
since the damping is due to very different mechanisms in theories with
or without additional hard thermal loop degrees of freedom.

As can be seen from the $\chi^2$-values reported above, the quality of
the fits shown in Figs.~\ref{fig:kappa} and \ref{fig:kappaprime} is
poor.  This is primarily due to the `pull' of the $\betaL=8.7$,
$m_D^2=8.2\, g^4T^2$ -point, which has very small statistical errors.
Exclusion of this point would make the fits acceptable; however, we do
not have any a priori reason for rejecting this point, so we keep it
in the analysis.  We take the badness of the fits into account by
expanding the errors in the quantities reported below by a factor
$\sqrt{\chi^2/\nu}$, where $\nu$ is the number of the degrees of
freedom in the fit.

In the limit $m_D^2\rightarrow\infty$ the coefficient of the ASY
scaling law becomes $\kappa' = 55.9 \pm 3.5$ using the polynomial
function in \eq\nr{fit1}.  However, more relevant for physical
applications is the Standard Model (MSM) value at $m_{\rm D}^2 =
(11/6) g^2 T^2$ and $\alpha_w = 1/30$.  This corresponds to point
$(g^2T/m_D)^2 = 0.23$ in \fig\ref{fig:kappaprime}, and thus there is
no need to extrapolate in $m_D^2$.

Finally, as discussed in the beginning of this section, the numerical
coefficient $0.68$ in \eq\nr{mdlatt} has an estimated (quite
conservative) systematic error bar $\pm 0.2$.  This error has little
effect on $\kappa'$ if we extrapolate to $m_D^2\rightarrow\infty$, but
at the physical MSM value it actually gives the leading contribution
to the total error.  When we take this into account, we obtain
the physical value 
\be
  \kappa'(\mbox{MSM}) = 46.6\pm 2.0_{\rm stat} \pm 3_{\rm syst}\,,
\ee
and the MSM Chern-Simons diffusion constant becomes (with combined
statistical and systematic errors)
\be
  \Gamma = 25.4 \pm 2.0 \alpha^5 T^4\,.
\ee
This value is in perfect agreement with the results obtained both with
the particle hard thermal loop degrees of freedom \cite{particles} and
with the classical Yang-Mills theory \cite{MooreRummukainen}.  

It is actually likely that the systematic error of the coefficient
$(0.68\pm 0.2)$ in \eq\nr{mdlatt} is overestimated: the mutual
consistency of the results obtained with and without the extra hard
thermal loop degrees of freedom becomes noticeably worse when this
coefficient is more than $\sim \pm 0.1$ away from the central value.

\section{Conclusions}
\label{Conclusion-sec}

Classical Yang-Mills theory plus hard thermal loops is the IR effective
theory for the SU(2) sector of the Standard Model above the electroweak
phase transition, and it should be used to determine the ``sphaleron
rate'' $\Gamma$, which sets the efficiency of baryon number violation.

We have developed a numerical implementation of the method of
auxiliary fields, originally developed by Blaizot and Iancu and by
Nair \cite{BlaizotIancu1,Nair:local,Blaizot:energy,Nair:hamiltonian}.
The auxiliary fields are expanded in spherical harmonics and the
series is truncated at a finite $\lmax$; then the theory is put on a
lattice.  The resulting numerical model is an efficient and
systematically improvable representation of the desired effective
theory.

Within errors we observe no lattice spacing dependence, and the
convergence to the large $\lmax$ limit is surprisingly rapid.  This
means that the lattice numerical model is both accurate and efficient.
Using it, we verify the Arnold-Son-Yaffe scaling behavior for $\Gamma$
\cite{ASY}, $\Gamma = \kappa' (g^2 T^2 / m_{\rm D}^2) \alpha^5 T^4.$
If we use the Standard Model values of $m_{\rm D}^2 = (11/6) g^2 T^2$
and $\alpha_w = 1/30$, the rate is
\beq
	\Gamma = (25.4 \pm 2.0) \alpha^5 T^4 ~\simeq ~ 
	(1.05 \pm 0.08) \times 10^{-6} T^4 \, .
\eeq
The final result is in good agreement with the results previously
obtained by Moore, Hu, and M\"{u}ller \cite{particles}.  It is also in
agreement with the results obtained in pure lattice Yang-Mills theory
\cite{MooreRummukainen} using the matching technique developed by Arnold
\cite{Arnoldlatt} to relate $\Gamma$ in pure classical lattice Yang-Mills
theory to its value in the quantum theory.

The problem of determining the sphaleron rate in Yang-Mills theory is
settled, at least at the $20 \%$ level.


\subsection*{Acknowledgments}

We would like to thank Jan Ambj{\o}rn, Peter Arnold,
Edmond Iancu, Arttu Rajantie, Dam Son and Larry Yaffe
for several very useful discussions.
This work was supported in part by the TMR network
``Finite temperature phase transitions in particle physics'', EU
contract no. ERBFMRXCT97-0122.  The computer simulations 
were performed on a Cray T3E at the Center for
Scientific Computing, Helsinki, Finland, and on an IBM SP2 at
UNI-C computing centre, Copenhagen, Denmark.

\renewcommand{\theequation}{\Alph{section}.\arabic{equation}}
\appendix

\section{Spherical coefficients $v_{mi}$ and $C_{lm,l'm',i}$}
\la{app:1}

In this appendix we give explicit expressions for the coefficients
$ C_{lm,l'm',i}$, \eq\nr{clm}, and $v_{mi}$, \eq\nr{dmi}.  
We use the conventional normalization for the spherical harmonic functions:
\be
 \int {d\Omega_v} Y^*_{lm} Y_{l'm'} = \delta_{l,l'}\delta_{m,m'}\,.
\ee
The coefficients $v_{mi}$ can be given simply as
\ba
 v_{mi} &\equiv&  \int {d\Omega_v} Y^*_{1m}(\vv) v_i \nonumber \\
        &=&\sqrt{\fr{4\pi}{6}}\delta_{i,1}(-\delta_{m,1}+\delta_{m,-1})
	+ i\sqrt{\fr{4\pi}{6}}\delta_{i,2}(\delta_{m,1}+\delta_{m,-1})
	+ \sqrt{\fr{4\pi}{3}} \delta_{i,3}\delta_{m,0}\,.
\la{vmi}
\ea
The matrix elements
\ba
  C_{lm,l'm',i} 
   &\equiv& \int {d\Omega_v} 
		Y^*_{lm}(\vv) v_i Y_{l'm'}(\vv) \nonumber \\
   &=& \sum_M v_{Mi} \int {d\Omega_v} 
		Y^*_{lm}(\vv) Y_{1M}(\vv) Y_{l'm'}(\vv)
\ea
are conveniently expressed in terms of the spherical components
\ba
  C_{lm,l'm',1} &=& \fr{1}{\sqrt{2}} (-C_{lm,l'm'}^+ + C_{lm,l'm'}^-) \nonumber \\
  C_{lm,l'm',2} &=& \fr{i}{\sqrt{2}} (C_{lm,l'm'}^+ + C_{lm,l'm'}^-) \nonumber \,.
\ea
Finally, we can write
\ba
C_{lm,l'm'}^+ &=& A(l',m') \delta_{l-1,l'}\delta_{m-1,m'}
	        - A(l,-m) \delta_{l+1,l'}\delta_{m-1,m'} \nonumber   \\
C_{lm,l'm'}^- &=& A(l',-m') \delta_{l-1,l'}\delta_{m+1,m'}
	        - A(l,m) \delta_{l+1,l'}\delta_{m+1,m'}    \la{clm-app} \\
C_{lm,l'm',3} &=&  B(l',m') \delta_{l-1,l'}\delta_{m,m'}
 		 + B(l,m)   \delta_{l+1,l'}\delta_{m,m'} \nonumber\,, 
\ea
where the coefficients are
\ba
  A(l,m) &=& \left[\fr{ (l+m+1)(l+m+2) }{ 2(2l+1)(2l+3) } \right]^{1/2}\,,\\
  B(l,m) &=& \left[\fr{ (l-m+1)(l+m+1) }{ (2l+1)(2l+3) } \right]^{1/2}\, .
\ea

\section{Linearized lattice propagator}
\la{app:doublers}

In this appendix we shall study the properties of the linearized gauge
field propagator with hard thermal loops on the lattice, as was done
in section \ref{Propagator-sec} in continuum.  As emphasized in
section~\ref{sec:lattice}, the second-order `leapfrog' update for
$W^a_{lm}$ decouples the even and odd parity sites from each other.
Here parity $p = \sum_i x_i + t/\dt$.  This decoupling creates extra
low-energy poles, {\em doublers\,}, in the gauge field propagator.
Here we show that these doublers are not relevant for the gauge field
dynamics.

Following Sect.~\ref{Propagator-sec} we linearize \eqs\nr{ustep},
\nr{estep} and \nr{wstep}, and study one $\vk$ mode
in isolation.  In general, this need not be parallel to any of the
major lattice axes.  In Sect.~\ref{sec:lattice} the spherical
harmonics were written in ``lattice basis'' coordinates, that is, the
$Y_{lm}$ components were mapped to lattice coordinate axis directions
in the customary way.  Naturally, there is no fundamental reason (only
great convenience) to do this, and here we choose to parametrize the
spherical functions as in Sect.~\ref{Propagator-sec}, so the $Y_{lm}$
``$x_3$-direction'' is parallel to $\vk$.  Then the transverse fields
should oscillate only along the plane defined by $m=\pm 1$ components.

Let us make a Fourier transformation of the lattice equations of
motion; after some work we obtain the
equations
\ba
  \widetilde \omega^2 A_m - \widetilde k^2 A_m &=& 
	\fr{m_{\rm D}^2}{\sqrt{3}} D_{mm'} \, W_{1m'}
\la{mom-a-latt} \\
  \hat \omega W_{lm} - \hat k_i \hat n_i 
	C_{lm,l'm',3} W_{l'm'} & = &
  \hat\omega \delta_{l,1} \fr{1}{\sqrt{3}} D_{mM}  A_M\,.
\la{mom-w-latt}
\ea
Here $\hat n = \vk/k$, and the lattice momentum functions are
\ba
  \widetilde k_i = \fr{2}{a} \sin \fr{k_i a}{2}, &~~& 
  \widetilde\omega = \fr{2}{\dt a}\sin \fr{\omega \dt a}{2},    \nonumber \\
  \hat k_i = \fr1{a} \sin k_i a, &~~& 
  \hat\omega = \fr1{\dt a}\sin \omega \dt a \,,
\ea 
and the matrix $D_{mm'}$, $m=0,\pm 1$, is defined as
\be
  D_{mm'}(\vk) = \sum_i \gamma_{mi}\, \cos\fr{k_i a}{2}\, 
	\gamma^*_{m'i} \h\h\h
  \gamma_{mi} = \sqrt{\fr{3}{4\pi}} R_{ij}(\hat n) v_{mj}
\ee
$R_{ij}$ is a rotation matrix which rotates $\hat n$ parallel to the
lattice $x_3$-axis, and $v_{mi}$ is defined in \nr{vmi}.  The matrix
$\gamma$ can be understood as a transformation between the lattice
coordinates and the $\hat n$-based spherical coordinates.  Here
$\gamma^\dagger\gamma = \gamma\gamma^\dagger = 1$.  The $\cos(k_i
a/2)$ factor in $D_{mm'}$ arises from the spatial symmetrization in
\eqs\nr{estep} and \nr{wstep}, and without this we would have $D=1$.
Also, due to the timelike symmetrization $\hat\omega$ instead of
$\widetilde\omega$ appears on the rhs of \eq\nr{mom-w-latt}.

Due to the matrix $D_{mm'}$ the equations of motion do not diagonalize
to independent $m$-components, as in the continuum.  However, if $\vk$
is parallel to any of the lattice axes we have 
\be
  D_{mm'} = \delta_{m,m'}\left(
	\delta_{|m|,1} + \delta_{m,0}\cos \fr{k a}{2}\right)\,.
\ee
In this case $D=1$ for transverse modes.  Now we can solve for
the transverse gauge field in \eqs\nr{mom-a-latt}, \nr{mom-w-latt}
as in Sect.~\ref{Propagator-sec}, and 
we obtain the lattice version of the inverse propagator
in \eq\nr{invprop}:
\be
   - \widetilde \omega^2 + \widetilde k^2 + \fr{m_{\rm D}^2}{3}
	\,\sum_{\alpha=1}^{\lmax}
   \fr{\hat\omega}{\hat\omega - \hat k \lambda^{\alpha}}
   (\xi_1^{\alpha})^2\, .
\la{invlattprop}
\ee
The pole structure of the propagator is not immediately evident from
\eq\nr{invlattprop}.  Nevertheless, the propagator has new doubler
poles for each physical low energy $(\omega\sim 0,k\sim 0)$ pole, as
shown in \fig\ref{fig:latprop}.  These are located in the corners of
the momentum square $0 \le k a \le \pi$, $0 \le \omega \dt a \le \pi$.

\begin{figure}[t]
\centerline{
\epsfysize=7.5cm\epsfbox{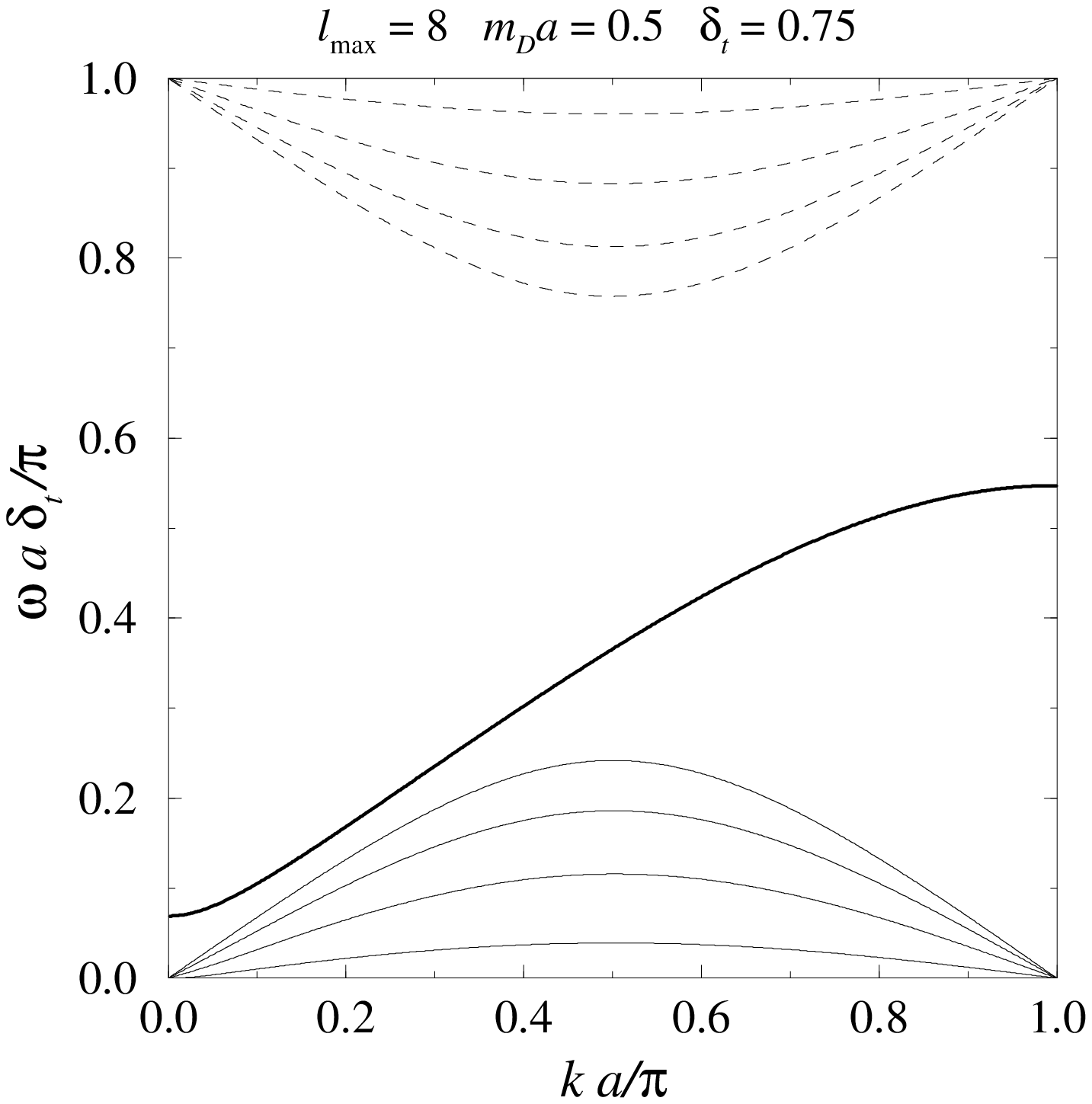}\hspace{4mm}
\epsfysize=7.5cm\epsfbox{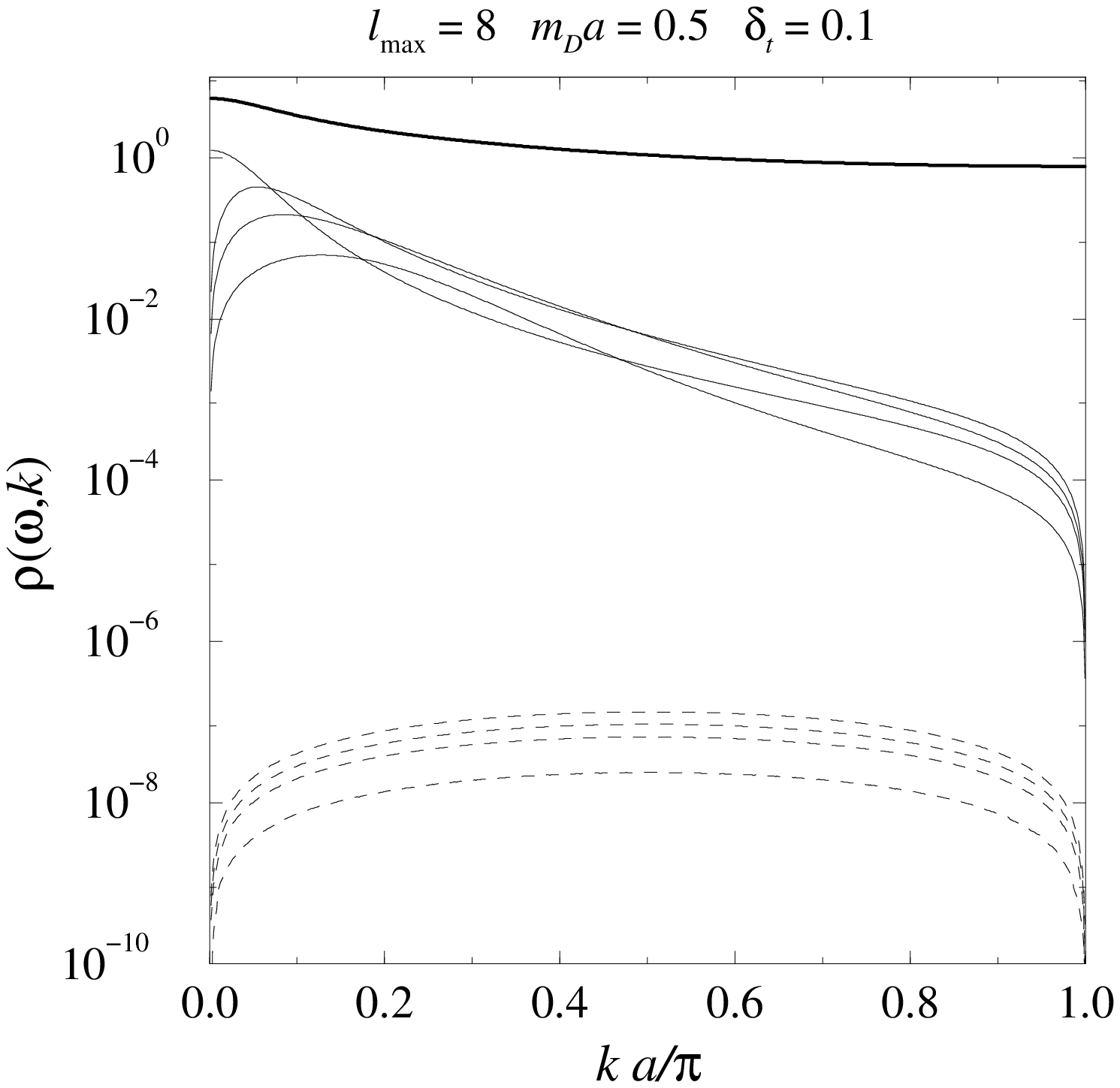}}
\caption[a]{{\em Left:} 
the positive frequency poles of the lattice gauge propagator within
the Brillouin zone, as functions of $k$, for $m_{\rm D} = 0.5/a$.  For each
of the continuum poles near origin there are doubler poles at the
corners of the Brillouin zone.  The thick line is the plasmon, and the
dashed lines are the timelike doublers.  For clarity, this figure is
plotted with unrealistically large $\dt = 0.75$. {\em Right:} The
spectral power (residue) of the poles when $\dt = 0.1$.  The lines are
as in the left panel.  None of the doublers carry a significant
fraction of the gauge field propagation.}
\la{fig:latprop}
\end{figure}
 
In \fig\ref{fig:latprop} we also show the spectral power 
\be
  \rho(\omega,k) = 
  \mbox{\,Im\,} \Delta (\omega+i\epsilon,\vk) 
\ee
of the poles (residue of the propagator).  At momenta close to the
lattice cutoff $\pi/a$ almost all of the power is carried by the 
plasmon pole, which does not have doublers.  Thus, at high momenta
the gauge field essentially decouples from $W$ fields, except for
a mass term which equals $m_{\rm D}/\sqrt{3}$ for $k$ along a lattice axis
(but not everywhere, it is zero at $a\vec{k} = (\pi,\pi,\pi)$).  
This also occurs in the continuum.
The poles other than plasmon are significant only around the
physical $k,\omega \sim 0$ corner.  Interestingly, even here
the plasmon has a power which is a factor of $\sim 5$ larger than
the other poles.  However, it is these poles which are significant
for the non-perturbative small-frequency physics.  

One might worry about the small-$k$ temporal doublers, which
correspond to modes which flip sign at each consecutive
timestep:  after all, these can have arbitarily
long spatial wavelength.  These are shown in \fig\ref{fig:latprop}
with dashed lines.  However, it turns out that these poles have even
much less power than the spatial doublers.  Thus, the existence of the
temporal poles should not affect the gauge field behavior at all.
Indeed, even in non-abelian theory simulations, where the fields
are fully interacting, we saw no significant
energy transfer between the temporal pole sector and the `normal' 
small frequency sector.

The propagator becomes much more complicated to study when we 
do not require that $\vk$ is along any of the lattice axes.  
However, the pole structure of the propagator is qualitatively
similar to any direction, and it will pick the full complement
of poles at each corner of the Brillouin zone.

In order to make the left panel of \fig\ref{fig:latprop} readable, it
is plotted using $\dt=0.75$.  This brings the frequency spread of the
poles to the same order of magnitude than the separation between the
temporal doublers and the other poles.  For a more realistic $\dt
\lsim 0.1$ the poles would lie almost along $\omega a \dt/\pi = 0,1$ lines.

\section{O(a) matching for $\Gamma$}
\label{Oamatch}

In this appendix we compute $O(a)$ radiative corrections which arise in
the infrared dynamics of the lattice theory due to the compact nature of
the gauge action and the manner in which the original equations were
discretized.  The goal is to find what modifications must be made to
$\Gamma$ and $m_{\rm D}^2$ (where here $m_{\rm D}^2$ is really being used to
represent the magnitude of the damping rate for gauge field
modes with $\omega \ll k \sim g^2T$, which determines the relevant
dynamics \cite{ASY,Arnoldlatt}; when we write $m_{\rm D}^2$ we mean the value
which gives the same damping rate using the continuum relation between
$m_{\rm D}^2$ and the damping rate). 

\subsection{Corrections to $t$ and $m_{\rm D}^2$}

To begin, we discuss the relation between the lattice and continuum
values for $E$, $D_j F_{ji}$ (meaning the first term on the right hand
side in 
Eq. (\ref{dE})), and time $t$.  Where possible we will suppress spatial
and group indices, in particular we refer to $D_j F_{ji}$ as $DF$.
We will write $E_L$ etc.~for the lattice
fields scaled to continuum units directly using
Eq. (\ref{latt_scaling}), but always using $a$ as given in
Eq. (\ref{betaL}).  The scaling between the continuum $A$ field and the
lattice one, defined as $U=\exp(igaA^a T^a)$, is gauge dependent, and we will
always use the continuum normalization.  The calculation here relies both
on \cite{Oapaper} and on Appendix A of \cite{Bodek_paper} very heavily.

Define the following renormalization constants:
\beqa
Z_g & = & \beta / \beta_L \simeq 1 - 0.61/\beta
	\, , \\
Z_E & = & \left[ 1 + \frac{N}{\beta} \left( \frac{1}{3} 
	\frac{\Sigma}{4 \pi} + 6 \frac{\xi}{4 \pi} \right) \right]
	\simeq 1 + .314 / \beta \, , \\
Z_W & = & 1 - \frac{N}{2 \beta} \frac{\Sigma}{4 \pi} \simeq
	1 - 0.2527 / \beta \, ,
\eeqa
where $Z_g$ is computed in \cite{Oapaper} and presented above in
Eq. (\ref{betaL}), $Z_E$ is first computed in the appendix of
\cite{Bodek_paper} (where it has the unfortunate notation of $(1+{\rm
corr})^2$), and $Z_W$ is new to this paper and discussed more in the
next subsection.

To begin observe that, just from its appearance in the Hamiltonian next
to $\beta_{\rm L}$, we have
\beq
\langle E_L^2 \rangle = Z_g \langle E^2_C \rangle \quad 
	\Rightarrow \quad E_L = Z_g^{1/2} E_C \, .
\eeq
Also, from \cite{Bodek_paper} Appendix A, we have 
\beq
\frac{dA_C}{dt_L} = Z_E^{1/2} E_L = Z_E^{1/2} Z_g^{1/2} E_C 
\quad \Rightarrow \quad
t_L = Z_E^{-1/2} Z_g^{-1/2} t_C \, .
\eeq
We apply this correction when we extract the continuum value of $\Gamma$
from data which appear as a time series in $t_L$, so $\Gamma$ quoted in
this paper is always scaled by $V t_C$ the continuum volume and time.
Next, to find the renormalization of $DF$, we can use pure gauge theory
relations
\beq
\frac{dE_L}{dt_L} = - DF_L \qquad {\rm and} \qquad
\frac{dE_C}{dt_C} = - DF_C \, ,
\eeq
to find that
\beq
DF_L = Z_E^{1/2} Z_g DF_C \, ,
\eeq
which we will need below.

\begin{figure}[t]
\centerline{\epsfxsize=5in\epsfbox{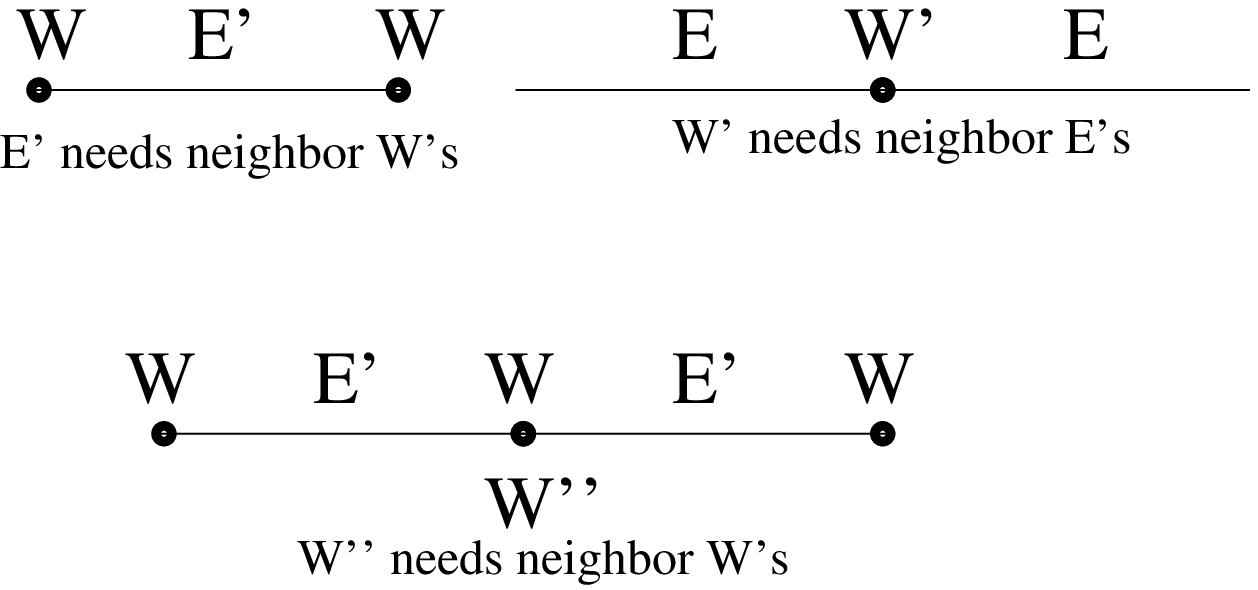}}
\caption{\label{appendixfig} Neighbor averaging involved in the updates
of $E$ due to $W$ and $W$ due to $E$.}
\end{figure}

To compute the radiative corrections in the $W$ field contribution to
the gauge field damping rate we need to consider the equations of motion
of the full system.  If there were only infrared fields, then the first
errors from our discretization (sampling neighbors to determine a
derivative , $\ldots$) would enter at $O(a^2)$, while here we will only
be interested in $O(a)$ effects.  Nevertheless, because of the different
behavior of UV modes on the lattice than in the continuum, 
three new corrections arise:  one in the $E$ field
source in the $W$ equation of motion, Eq. (\ref{dW}); 
one in the $W$ field source in the Yang-Mills-Maxwell-Ampere equation,
Eq. (\ref{dE}); and one in the $W$ field convective covariant derivative 
in Eq. (\ref{dW}).  The former two occur because, whereas in the
continuum these equations relate fields at the same point
(see Eqs. (\ref{update-lm}) and (\ref{current-lm})), on the lattice they
involve averages over nearby points, see Fig. \ref{appendixfig}.  
The latter correction occurs
because the derivative term necessarily involves the gauge field
connection.  In each case the gauge field connection enters, and the
interaction receives tadpole contributions which are absent in the
continuum.  As a result, the effective IR equations of motion look (in a
simplified notation, dropping all subscripts including $l,m$ indices
and all Clebsch-Gordan
coefficients, which are the same as in the text of the paper)
\beqa
\label{app_E_update}
\frac{dE_L}{dt_L} & = & - DF_L - m_{\rm D}^2
	\kappa_1 W_L \, , \\
\label{app_W_update}
\frac{dW_L}{dt_L} & = & \kappa_2 E_L
	- \kappa_3 v \cdot D_C W_L \, , \\
\kappa_1 \kappa_2 & = & Z_W \, , \qquad \kappa_3 = Z^2_W \, .
\eeqa
Here $m_{\rm D}^2$ is the lattice $m_{\rm D}^2$ parameter converted to physical
units by scaling with factors of $a$.
We derive the size of the corrections $\kappa_{1,2,3}$
in the next subsection.

Re-arranging Eq. (\ref{app_W_update}) to
\beq
\left[ \kappa_3^{-1} \frac{d}{dt_L} + v \cdot D_C \right] W = 
	\kappa_3^{-1} \kappa_2 E_L \, ,
\eeq
formally inverting it, and substituting the solution for $W$ into
Eq. (\ref{app_E_update}), gives
\beq
\frac{dE_L}{dt_L} = - DF_L - m_{\rm D}^2 \left[ \kappa_3^{-1} 
	\frac{d}{dt_L} + v \cdot D_C \right]^{-1} \kappa_1 
	\kappa_2 \kappa_3^{-1} E_L \, .
\eeq
It has been argued in \cite{HuetSon,Son} that in the overdamped case it
is permissible to drop both the $dE/dt$ term and the time derivative
appearing in the inverse operator.  Technically doing so commits an
error of order $O(g^4 T^2 / m_{\rm D}^2)$.  Note however that errors of
precisely this size already arise from subleading corrections to the
hard classical lattice mode contribution in Eq. (\ref{mdlatt}).
Therefore, in Figure \ref{fig:kappaprime} there is an unknown systematic
error in the slope of the fit line, which we will not be able to
eliminate.  However, we can still ask to make all $O(a)$ corrections
which would affect the intercept.  To do so we are permitted to drop the
time derivatives mentioned above, giving 
\beq
m_{\rm D}^2 \left( Z_W^{-1} Z_E^{-1/2} Z_g^{-1/2} \right)
	\left[ v \cdot D_C \right]^{-1} \frac{dA_C}{dt_C} = DF_C \, ,
\eeq
which gives us the $O(a)$ renormalization appropriate for the $W$ field
Debye mass term, namely, the value to use as the strength of the gauge
field damping term is
\beq
Z_{m_D}^{-1} m_{\rm D}^2 \equiv 
Z_W^{-1} Z_E^{-1/2} Z_g^{-1/2} \: m_{\rm D}^2 \, .
\la{Zmd}
\eeq

\subsection{Evaluating $\kappa_{1,2,3}$}

We see from Fig. \ref{appendixfig} that the product $\kappa_1 \kappa_2$
arises from the difference between $W^a(x)$ and 
$(1/4)[ \trans^{ab}_i W^b(x+ia) + 2 W^a(x) + \trans^{ab}_{-i}
W^b(x-ia)]$.  We compute this difference for a very slowly varying $W$
field in Coulomb gauge.  (In this gauge the effects of the $A_0$ field on
the dynamics do not differ between the lattice and continuum, see
\cite{Bodek_paper}.)

The parallel transport $\trans^{ab}_i W^b(x+ia)$ is
\beqa
T^c(U_i(x) W_+ U^{\dagger}_i(x))^c & = & \left( 1 + iga T^a A_i^a(x)
- \frac{g^2a^2}{2} T^aT^b A^a_i(x)A^b_i(x) + \ldots \right) \nonumber \\
	& &  \qquad \times 
 T^cW^c \times \bigg(
	{\rm same}, \; i \leftrightarrow -i \bigg) \, ,
\eeqa
and on expanding (and writing $(1/4) (W(x+ia) + 2 W(x) + W(x-ia) )$ as
$W$) eventually gives
\beqa
&&\frac{\trans^{ab}_i W^b(x+ia) + 2 W^a(x) + \trans^{ab}_{-i}
W^b(x-ia)}{4} \nonumber \\
& = & W^aT^a + \frac{ga}{4}W^b
	\left(A^c_i(x)-A^c_i(x-ia)\right) f_{abc}T^a \nonumber \\ & & 
	+\frac{g^2a^2}{8}W^a \left(A^b_i(x)A^c_i(x) + A^b_i(x-ia)
	A^c_i(x-ia) \right) [T^c,[T^a,T^b]] + O(a^3) \, .
\eeqa
In momentum space the first term here is $-f^{abc}(ga^2/4)\int_l \tilde{l}_i
W^b(l-k)A^c_i(-l)$.  It cannot lead to a contribution proportional to
$W(k)$ because $\langle f^{abc} W^a(k) W^b(l-k) A^c(-l) \rangle = 0$.
Therefore the first term does not rescale the interaction, so it does
not contribute to $\kappa_1 \kappa_2$.
However, the last term does lead to renormalization of $W^a T^a$, of
magnitude 
\beq
f_{adb} f_{bde} T^e \frac{g^2 aT}{4} \int \frac{d^3 (ak)}{(2 \pi)^3}
\frac{1}{a^2\tilde{k}^2} \left( 1 - \frac{\tilde{k}_1^2}{\tilde{k}^2} 
	\right) \cos^2 (ak_1/2) \, ,
\eeq
where $\tilde{k} \equiv (2/a)\sin(ak/2)$ and the integral runs over the
Brillioun zone, $ak_i \in [- \pi , \pi]$.  Using identities from
\cite{Oapaper}, the value of the integral is $(1/2) \Sigma/(4 \pi)$.
Therefore the final rescaling we find is
\beq
\kappa_1 \kappa_2 \equiv Z_W = 1 - \frac{N}{2 \beta} 
	\frac{\Sigma}{4 \pi} \, .
\eeq

The calculation of $\kappa_3$ proceeds similarly.  Here we need to
compute $(\trans^{ab}_i W^b(x+ia) - \trans^{ab}_{-i}W^b(x-ia))/2$.  
The linear in $A$ term now contains $A_i(x) + A_i(x-ia)$, and just gives
the $A$ field part of the continuum $D^{ab}_i = \delta^{ab} \partial_i -
gf_{abc} A^c_i$.  The quadratic in $A$ terms perform the renormalization
of $\partial_i$ and give exactly twice the corresponding contribution to
$\kappa_1 \kappa_2$, 
because in that case only half of the expression arose from $W$ fields
which are parallel transported.  Hence we find $\kappa_3 = Z_W^2$.

\newcommand{\hep}[1]{[{#1}]}
\newcommand{\heplat}[1]{[{hep-lat/#1}]}
\newcommand{\hepth}[1]{[{hep-th/#1}]}

\end{document}